\documentclass[final]{elsart}

\usepackage[dvips]{graphicx}
\usepackage{amssymb}
\usepackage{amsmath}
\usepackage{subfigure}
\usepackage{color}

\begin{document}

\begin{frontmatter}

\title{\Large\bf Energy determination of cosmic ray showers in surface arrays using
signal inference at a single distance from the core}

\author[ALCALA]{G. Ros\corauthref{cor1}},
\author[UNAM]{G. Medina-Tanco},
\author[ALCALA]{L. del Peral},
\author[UNAM]{J. C. D'Olivo},
\author[UCM]{F. Arqueros},
\author[ALCALA]{M. D. Rodr\'{\i}guez-Fr\'{\i}as}.

\address[ALCALA]{Space Plasmas and AStroparticle Group, Dpto. F\'isica, Universidad de Alcal\'a 
Ctra. Madrid-Barcelona km. 33. Alcal\'a de Henares, E-28871 (Spain).} 
\address[UNAM]{Instituto de Ciencias Nucleares, UNAM, Circuito Exteriror S/N, Ciudad Universitaria,
M\'exico D. F. 04510, M\'exico.}
\address[UCM]{Dpto. F\'{\i}sica At\'omica, Molecular y Nuclear, Facultad de
F\'{\i}sica, Universidad Complutense de Madrid, Ciudad Universitaria 28040, Madrid (Spain).}

\corauth[cor1]{e-mail: german.ros@uah.es. In collaboration with
Universidad Complutense de Madrid and Universidad Nacional
Aut\'onoma de M\'exico.}

\begin{abstract}

In most high energy cosmic ray surface arrays, the primary energy is currently determined
from the value of the lateral distribution function at a fixed distance from the shower core, 
$r_{0}$. The value of $r_{0}$ is mainly related to the geometry
of the array and is, therefore, considered as fixed independently of the shower
energy or direction. We argue, however, that the dependence of $r_{0}$ on energy
and zenith angle is not negligible. Therefore, in the present work we propose a new 
characteristic distance, which we call $r_{opt}$, specifically determined for each individual shower, 
with the objective of optimizing the energy reconstruction. 
This parameter may not only improve the energy determination, but also allow a more reliable 
reconstruction of the shape and position of rapidly varying spectral features. 
We show that the use of a specific $r_{opt}$ determined on a shower-to-shower basis, instead of using a fixed characteristic value, is of particular benefit in dealing with the energy reconstruction of events
with saturated detectors, which are in general a large fraction of all the events detected by an
array as energy increases. Furthermore, the $r_{opt}$ approach has the additional advantage of 
applying the same unified treatment for all detected events, regardless of whether they have saturated detectors or not.

\end{abstract}


\end{frontmatter}

\section{Introduction}

The energy spectrum of cosmic rays (CR) is observed from energies below 1 GeV up 
to more than $10^{20}$ eV. CR hit the top of the atmosphere at a rate  
$\gtrsim 10^{3}$ per square meter per second. For energies below  $\sim 10^{2}$ TeV, CR 
can be detected by direct measurement from high altitude balloons or satellites, albeit with 
rapidly decreasing statistics since the CR spectrum is very steep, $\sim E^{-2.7}$. For $E 
\gtrsim 10^{2}$ TeV, the point of maximum development of the
cascades of daughter particles initiated by collisions with atmospheric nuclei starts reaching the 
ground at high altitudes. From that point onward, the properties of primary cosmic ray can be 
determined indirectly from the measurement of extensive air showers (EAS).

Two different experimental approaches have been traditionally used to study the highest energy EAS. 
The first one consists on the inference of lateral distribution of particles
from the discrete sampling of the shower front at ground level by a surface array of
detectors. Scintillators (e.g., Volcano Ranch, AGASA, KASCADE) and water Cherenkov 
detectors (e.g., Haverah Park) have been mainly used for this purpose. The second technique
consists on the reconstruction of the longitudinal profile of the shower from the 
fluorescence light produced by atmospheric Nitrogen as it is excited by charged EAS 
particles along the atmosphere (e.g., Fly's Eye, HiRes). The latter is considered to 
be close to a calorimetric measurement of the primary CR particle energy. Extensive 
reviews on CR theory and experiments can be found in \cite{Nagano,HillasReview,Unger}.

A special case from the experimental point of view is the Pierre Auger Observatory \cite{Auger} 
which pioneers the simultaneous use of both techniques, water Cherenkov detectors and
fluorescence telescopes. For these {\it hybrid} events, systematic errors in their energy estimate are greatly reduced. Unfortunately, fluorescence can only be observed during dark nights and, consequently, this technique can only be applied to $\sim 13$\% of the incoming events. Therefore, even if the hybrid 
technique, when simultaneously available, allows for an independent calibration of a ground detector, 
high statistics must come from surface arrays with a duty cycle close to 100\%.

In this work we are interested on the highest energies, $E > 10^{17}$ eV, at which point extragalactic
flux is likely to penetrate the Galaxy and start dominating the CR flux. At these energies at least three very important spectral features are located: the second knee, the ankle and the GZK complex (bump plus steep flux suppression) \cite{GMTancoPeVZeV}. The determination of their exact position and shape is a fundamental experimental problem. Therefore, the most relevant parameter in this spectral region is, arguably, the primary energy of the impinging CR.  

The procedure to determine the primary energy in sufrace arrays is a two step process. 
First, the lateral distribution function (LDF), i.e. the shower particle 
density or signal  versus distance to the shower axis, is fitted assuming a known functional form. 
This fit suffers from uncertainties related to the statistical shower fluctuations, 
the uncertainties in core location and the ignorance of the exact form of the LDF. The normalization 
constant of the LDF of an extensive air shower is a monotonous (almost linear) increasing function 
of the energy of the primary cosmic ray. Therefore, 
Hillas \cite{Hillas} proposed to use the interpolated signal at some 
fixed, {\it characteristic distance\/} from the shower core, $S(r_{0})$, at which fluctuations in the LDF are minimal. 
The uncertainty due to the lack of knowledge of the LDF is also minimized by this procedure \cite{AugerLDF}.  The use of the signal interpolated at $r_{0}$, $S(r_0)$ is widely used as energy estimator by surface detector arrays. AGASA \cite{AGASALDF,AGASADai}, Yakutsk \cite{Yakutsk_ropt} and Haverah Park \cite{HP_ropt}, for example, choose $r_{0}=600\; m$, while Auger uses 1000 m due to its larger array spacing \cite{AugerSpectrumPRL}. 
The characteristic distance $r_{0}$ is mainly, although not completely, determined by the geometry of the array. Thus, the same value of $r_{0}$ is used to estimate the energy for all the showers, regardless of primary energy or incoming direction. In the second step, there are at least two possible approaches 
to calibrate $S(r_0)$ as a function of primary energy: either via Monte Carlo simulations or, as in the case of Auger, by using the almost calorimetric measurement obtained from the fluorescence observation of 
high quality hybrid events \cite{AugerSpectrumPRL}.

As an alternative, but motivated by Hillas' original idea \cite{Hillas}, in the present work we focus 
in the shower-to-shower determination of an {\it optimal} distance to the core, which we name 
hereafter $r_{opt}$, at which the interpolation of the {\it reconstructed} signal is the best energy 
estimator for each individual shower, regardless of whether this point is actually 
the one that minimizes shower to shower LDF fluctuations. 

We perform a detailed study of $r_{opt}$ as function of array spacing, primary energy and 
the zenith angle of the incoming cosmic ray and demonstrate that, although array geometry is an 
important underlying factor, the dependence of $r_{opt}$ on the remaining parameters is not 
negligible. We study the bias associated with both techniques, $r_{0}$ and $r_{opt}$, and show
that, if the dynamical range of the detector covers a wide interval of energies, it is much
safer to estimate an $r_{opt}$ for the energy reconstruction of each individual event than to 
fix a single $r_{0}$ for the whole data set. In fact, not only the bias as a function of energy can
be kept negligible over at least 2.5 decades in energy, but also the error distribution functions
are much better behaved, i,e, without appreciable kurtosis or skewness and very much 
Gaussian in the mentioned energy range. The latter has a potential impact in the reconstruction 
accuracy of the energy spectrum. We demonstrate this by applying a fixed $r_{0}$ as well as 
a shower-to-shower $r_{opt}$, to a simplified version of the actual energy spectrum between 
$\sim 1$ and $\sim 100$ EeV.

A further advantage of the $r_{opt}$ approach is the straightforward treatment of events with 
saturated detectors. The problem of saturation is very common in all surface experiments, 
specially when dealing with high energy vertical showers. In fact, at the highest energies 
inside the designed dynamical range of any experiment, usually events with saturated detectors 
can account for a large, if not dominant, fraction of all the observed events. Different strategies 
have been used to deal with them. In some cases saturated detectors are directly discarded 
from the LDF fit, while in others the saturation value is used as a lower limit to the true signal 
during the fitting procedure. The Auger Collaboration is developing at present special, more
sophisticated algorithms to estimate the signal of a saturated detector \cite{MarisThesis} in
order to more properly account for them in the LDF fit. We show here that it is actually not
possible to define a single characteristic $r_0$ distance for both kinds of events. In fact, even 
if well defined medians values of $r_{opt}$ for events with and without 
saturated detectors do exist, the dispersions around the median at any energy are so large that both sets cannot be clearly differentiated as to use, for example, just two fixed distances instead of a single one. Nevertheless, using a shower-to-shower $r_{opt}$ distance, the inferred energy is unbiased for events with and without saturated detectors alike. 
This reconstruction strategy allows for an homogeneous treatment of the data set regardless of the increasing number events with saturated detectors when the energy increases.


In a recent work \cite{Newton}, Newton and co-workers also estimated an optimal shower-to-shower
distance, but used a different algorithm and with a somewhat different scope. They were mainly 
concerned with demonstrating the existence of a single distance for any given shower at which 
fluctuations in the LDF are minimum. By assuming that such fluctuations can be well described 
by the fluctuations of just one parameter, the slope of the LDF, externally fed into their 
procedure, and using a combination of simulations and semi-analytical analysis, they claim that, 
regardless of the functional form of the LDF considered, there exist a {\it convergence} point 
of the LDFs, at a characteristic distance they call optimal, where shower-to-shower fluctuations 
are minimal. Their results, combined together for a mix of energies drawn from a flat spectrum, 
seem to support their claim and lead them to the conclusion that a single fixed distance, depending
only on the geometry and spacing of a given array, would be a good choice for the energy determination 
in the whole energy range of the experiment. Furthermore, it is not clear from their study how to deal 
with the events with saturated detectors in the later scenario.

Alternatively, in the present work we do not constrain the parameters of the LDF, which are an output of the fit to the simulated data. We introduce instead reliable error estimations
for the reconstruction of the core position, as calculated by \cite{M.C.Medina} for arrays of varying spacing as a function of energy. 
Furthermore, our final scope is the determination of energy all along the dynamical range of an
experiment, and not the study of the manifestation of signal fluctuations in the LDF. 
Therefore, we study in detail the dependence of $r_{opt}$ and of its distribution function as a function of energy, zenith angle and array spacing. This study is performed for events with and without saturated detectors. We also give a comparative description of error and biases for the fix distance and the $r_{opt}$ distance approaches in that parameter space. In the 
same line, we further extend our analysis to the reconstruction of a simulated energy spectrum of
known shape, and show what the potential effects are of using each technique.

The paper is organized as follows. Section 2 describes our general algorithm. Two different detector arrays are considered, scintillators and water Cherenkov tanks. In Section 3, in order to study the $r_{opt}$ dependencies with array spacing and the energy and incoming direction of primary cosmic ray, water Cherenkov (Auger-like) stations have been used. In Section 4 we deal with the issue of energy determination.
In that analysis, (AGASA-like) scintillators are considered. A general discussion and conclusions are given in Section 5. While different detectors are used in Sections 3 and 4, the algorithm to find $r_{opt}$ is the same for both and the results and conclusions of the paper are not affected by the array under consideration.

\section{Algorithm}\label{Algorithm}

The basic idea of our algorithm is to estimate the optimum distance to the
core at which to determine the energy of a shower under the most realistic 
possible conditions.

We assume a certain analytical LDF as the intrinsic average lateral distribution of 
particles inside the shower front as a function of the distance to the core. For the chosen 
energy and geometry of the event (azimuth, zenith and true core position), this function is 
used to estimate the average LDF value at the actual position of each detector. Afterward,
a signal is calculated using the previous average value as the mean of a Poissonian distribution. 
If the calculated signal falls between a minimum threshold and an upper limit corresponding to a saturation condition, it is assigned to the detector. 
In the case of saturation, the event is kept, but with a flag indicating this fact, although
the saturated detectors are not used in the subsequent analysis, i.e. in fitting the LDF. 
Once a set of triggered detectors participating in the event has been defined, the 
reconstructed LDF is emulated by fitting an {\it experimental\/} LDF, which depends 
on the detector array under consideration, and is not necessarily the {\it real\/} LDF used in 
the first step to generate the event. The LDF fit requires an estimate of the core 
position. Such estimate is an important component of the analysis of the event 
and comes, in practice, from a global reconstruction procedure which implies an 
energy dependent error in the inferred position of the core. In our algorithm, we
simulate this error by shifting the reconstructed core position according to its 
experimentally determined Gaussian distribution function. 
For each {\it shifted\/} core position an independent
LDF fit is performed. We define the optimum distance to the core $r_{opt}$ as the interpolated
distance at which the dispersion between the several LDF fits is minimal. 
We argue that the interpolated signal at this point is the optimum estimator of the energy 
of a real event and constitutes the operational definition of our parameter $r_{opt}$ \cite{GustavoAlgorithm}. 

We use the following numerical approach to simulate EAS detection in a surface 
array. The array is a set of equally spaced detectors, located at the vertices
of an infinite grid of triangular elementary cells with variable spacing. The input parameters 
of an event are its energy, azimuthal and zenith angles and core position. The identity of the primary particle is not taken into account since differences in composition produce only a
small effect in the error distribution function of the reconstructed core position \cite{M.C.Medina}
which, in turn, when combined with the use of an experimental LDF maps into a negligible variation in both $r_{0}$ and $r_{opt}$. 

Whenever we simulate a water Cherenkov detector, we assume that 
the true lateral distribution of the signal is best represented by a Nishimura-Kamata-Greisen
(NKG) function. This functional form was first obtained in an analytical study of the lateral development of electromagnetic showers in \cite{NKG}, and later extended to the hadronic initiated showers because the electromagnetic particles represent around the 90\% of the total particles of the shower. The NKG selected is normalized at 1 km in the same way as the reported by Auger in \cite{Auger}:

\begin{equation}\label{AugerLDF}
S(r,E,\theta)=\frac{7.53\;E^{0.95}\;2^{\beta(\theta)}}{\sqrt{1+11.8[sec(\theta)-1]^2}}\times
r^{-\beta(\theta)}\times(1+r)^{-\beta(\theta)}
\end{equation}

\noindent where $r$ is the distance to the shower axis expressed in km, $E$ is the energy of the primary in EeV, $\theta$ is the zenith angle and  $\beta(\theta)=3.1-0.7sec(\theta)$. The signal in eq. \ref{AugerLDF} is expressed 
in vertical equivalent muons (VEM), which correspond to the signal deposited by one vertical
muon in an Auger water Cherenkov tank.

We use eq. \ref{AugerLDF} as the real LDF to simulate any given incoming 
event. The measured signal at each station is obtained with a Poissonian probability 
distribution function whose mean is given by eq. \ref{AugerLDF}, the ''true'' LDF.
The trigger condition is set to $S(r) = 3.0$ VEM. The saturation value is fixed at
S($0.2$ km, $1$ EeV, $0^o$). These values are compatible with the equivalent Auger parameters.

The uncertainty in core determination depends on the array geometry and primary
energy and it has been estimated for a variety of cases in reference \cite{M.C.Medina}.
We simulate the reconstruction uncertainty of the core using a Gaussian distribution function
centered at the position of the real core, with standard deviation given by \cite{M.C.Medina}
as a function of the energy of the shower for the array spacing under consideration.

For any shower, the following procedure is used to obtain the optimum 
distance $r_{opt}$. Throughout the procedure, we try to mimic, as far as 
possible, the actual reconstruction procedure. As explained earlier, 
several fits to the LDF are performed for any event, each one with its own 
estimated core position. Since the exact functional form of the LDF function is not crucial 
\cite{AugerLDF} we use a generic LDF parametrization to fit the signals of the triggered stations:

\begin{equation}\label{CherenkovTanks_LDFFit}
\log S(r)=a_{1}r^{-a_{2}}+a_{3}
\end{equation}

The uncertainty in the core position used for each one of the LDF fits corresponding 
to a given event, is accounted for by randomly shifting that point 50 times with the same Gaussian 
probability distribution function referred above centered at the position of the 
reconstructed core.
 
For each new core position, the LDF fit is performed using eq. \ref{CherenkovTanks_LDFFit}. 
The slope and the normalization constant of each LDF are determined from the fitting procedure. 
The $r_{opt}$ value is defined as the point at which the dispersion among the interpolated 
signals over the several LDFs goes through a minimum. 

Therefore, the implementation of the algorithm requires a two step
process: first, a global fit to the LDF is performed in order to get an estimate of 
the reconstructed core position and, second, the reconstructed core itself is 
fluctuated and $r_{opt}$ obtained using the procedure explained in the previous 
paragraph.

When simulating scintillators as those of the AGASA experiment, we follow exactly the same
procedure as before, but we use the following LDF \cite{AGASASpectrum} instead of 
eq. \ref{AugerLDF}:

\begin{eqnarray}\label{AgasaLDF}
\rho(r, E, \theta) &=& 49.676 \times 7.55^{\eta(\theta)-1.2} \times   f_s(\theta) \times
                 E^{1/1.03} \times \\ \nonumber
         & &  \left(\frac{r}{r_{M}}\right)^{-1.2} 
\left(1+\frac{r}{r_{M}}\right)^{-(\eta(\theta)-1.2)} 
\left(1+ r^{2}\right)^{-0.6}
\end{eqnarray}

\noindent where $\rho$ is given in m$^{-2}$, distances are in m, $r_{M}=91.6\; m$ is the Moliere radius at AGASA altitude, $\eta(\theta)=3.84-2.15(\sec(\theta)-1)$ and $f_s(\theta)$ is the attenuation curve:

\begin{eqnarray}\label{Agasa_Att_Curve}
f_{s}(\theta)  & = & exp\left[-\frac{X_0}{\Lambda_1}(sec\theta-1)-\frac{X_0}{\Lambda_2}(sec\,\theta-1)^2\right]
\end{eqnarray}

\noindent where $X_0=920\; g/cm^2$, $\Lambda_1=500\; g/cm^2$ and $\Lambda_2=594\; g/cm^2$ for showers with $\theta \leq 45^o$. The signal is fluctuated, as always, with a Poissonian distribution.

In this case, the trigger condition is selected in such a way that the signal is not
dominated by fluctuations. In particular, we use a vale of $\rho$ such that fluctuations
account at most by $50$\% of the signal. The saturation value is $\rho(0.2$ km $,1$ EeV $,0^o)$. 

In the AGASA case, we use as shower generator eq. \ref{AgasaLDF}, and perform
the subsequent fitting procedure using an LDF with the functional form ``observed'' by AGASA:

\begin{equation}\label{Scintillators_LDFFit}
\log \rho(r)=a_{1}-a_{2}\log(r/r_{M})-0.6\log(1+ \left(r/ 1000
\mbox{m} \right)^{2})
\end{equation}

\noindent which is formally equivalent to eq. \ref{AgasaLDF} for $r >> r_{M}$ \cite{AGASALDF}.

\section{$r_{opt}$ dependence on array spacing, energy and zenith angle.}

We consider in this section water Cherenkov detectors with separations of 433, 750, 866 and 1500 m,
as well as primary energies varying from $10^{17}$ to $10^{19.5}$ eV. We 
use eq. \ref{AugerLDF} to generate the signals and eq. \ref{CherenkovTanks_LDFFit} to fit the LDF. 

In all cases we consider a uniform distribution in azimuth and zenith angles as explained 
in each figure. Shower cores are uniformly distributed inside an elementary cell of the array.

Events with and without saturated detectors lead frequently to systematically different behaviors 
regarding the relationship between $r_{0}$ and $r_{opt}$ under discussion and, in principle, 
should be treated differently during data processing. Thus, in what follows, we will analyze them 
separately whenever appropriate.

\begin{figure}[!bt]
  \centerline{
  \subfigure{\includegraphics[width=7cm]{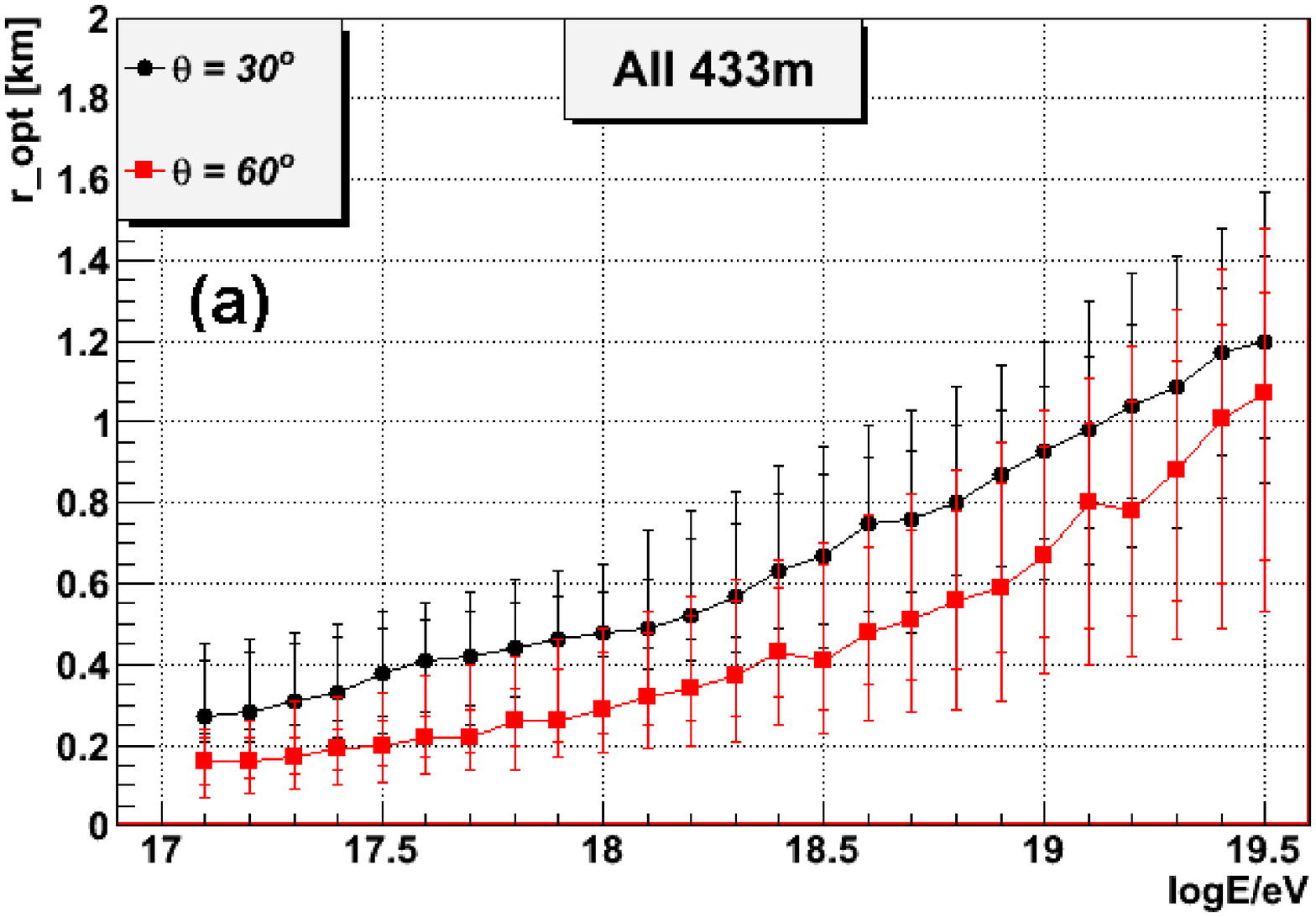}}
  \hfil
  \subfigure{\includegraphics[width=7cm]{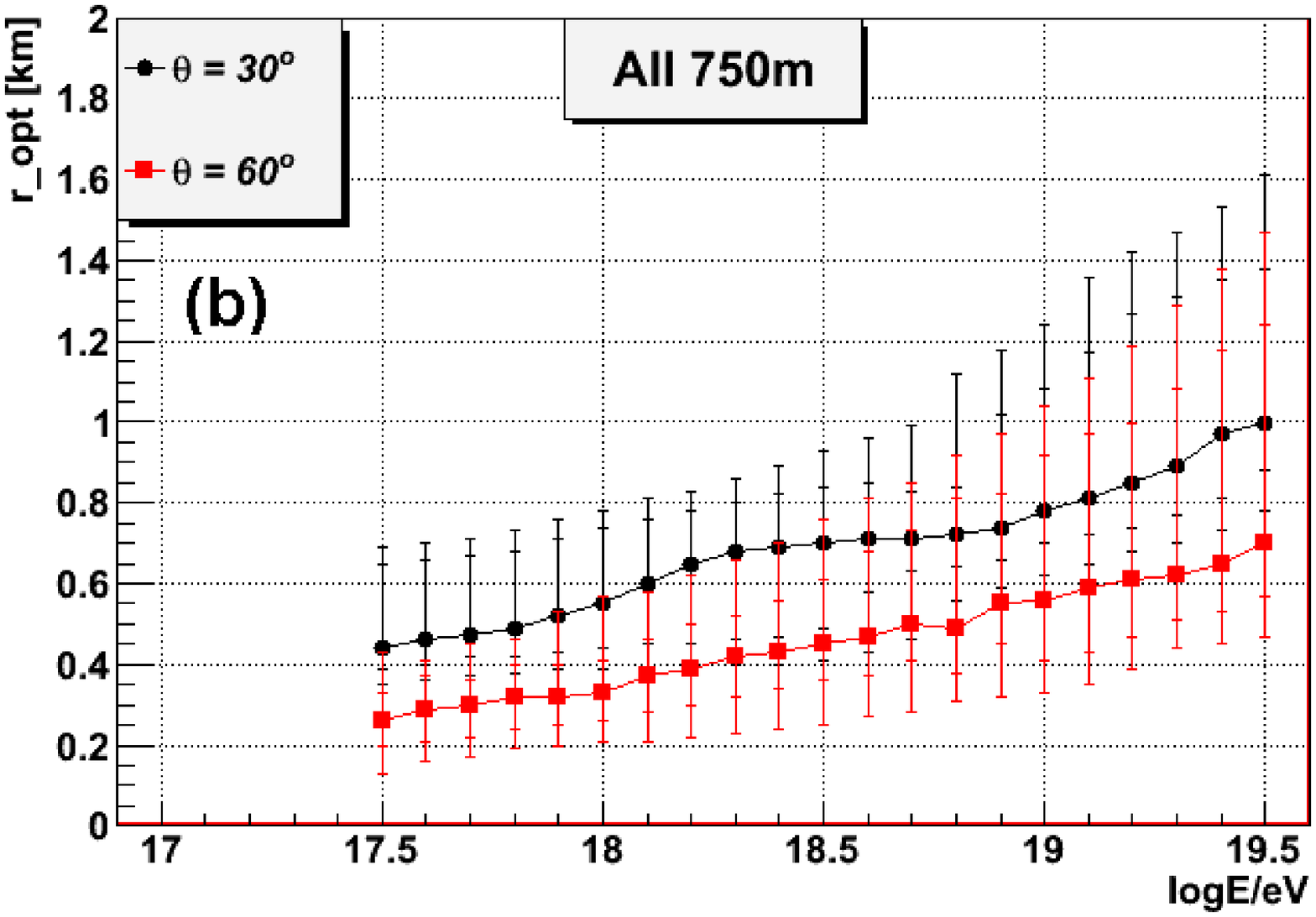}}
  }
  \centerline{
  \subfigure{\includegraphics[width=7cm]{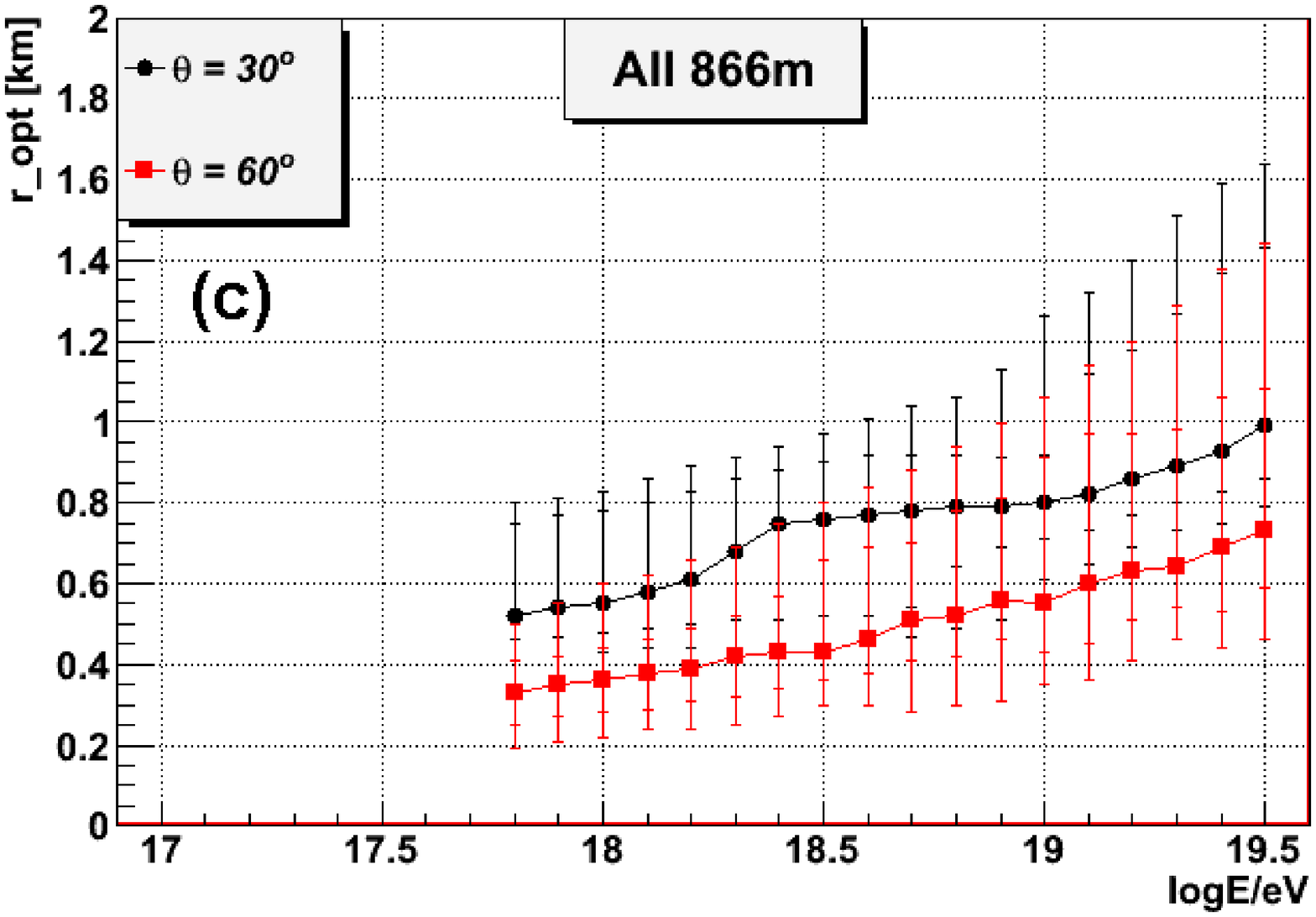}}
  \hfil
  \subfigure{\includegraphics[width=7cm]{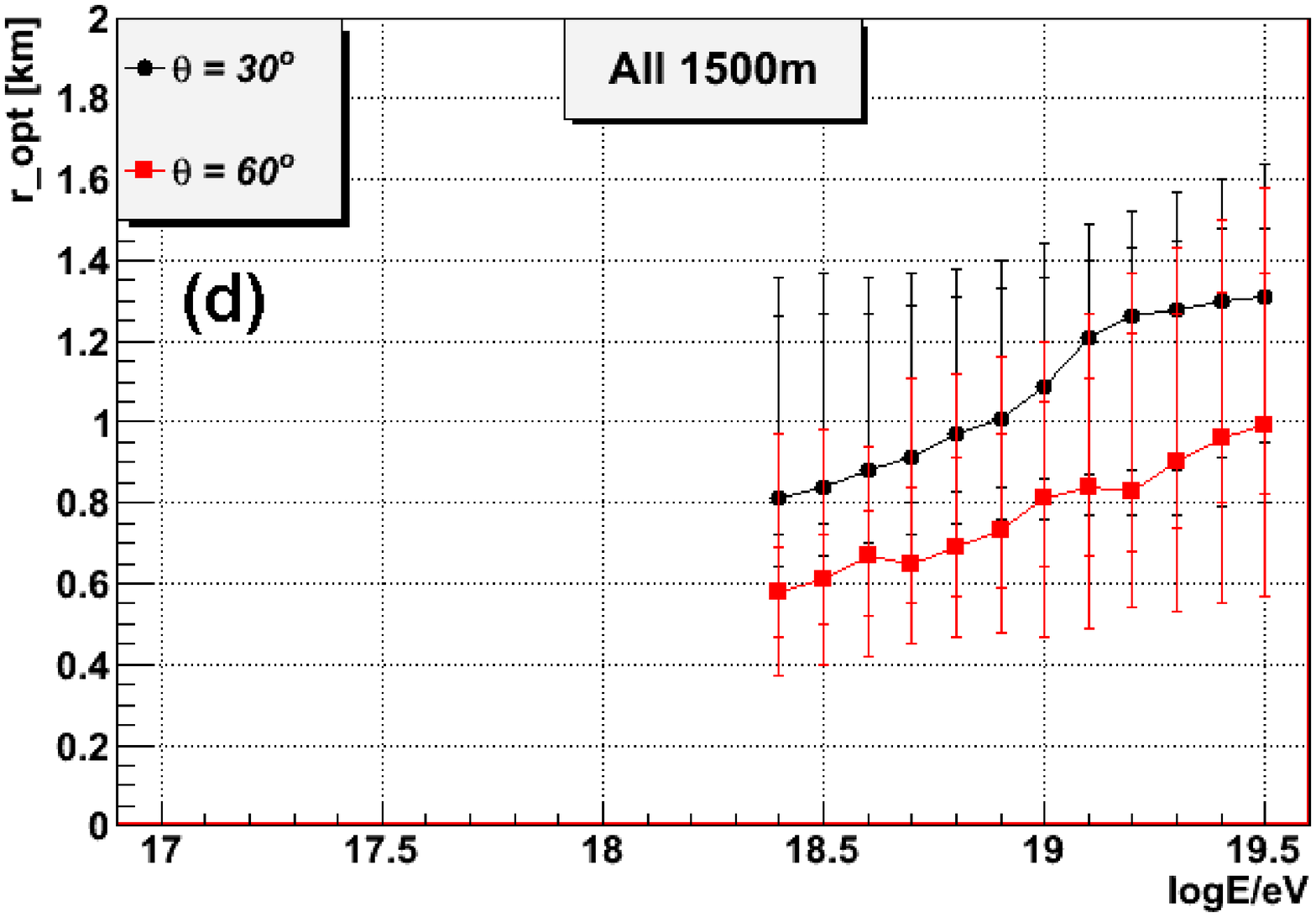}}
  }
  \caption{$r_{opt}$ vs. energy for different array spacing and zenith angle. The error bars represent the 68\% and 95\% C.L. The labell $All$ means that events with and without saturated detectors are both included.}
  \label{fig:ropt_vs_energy_all}
\end{figure}

\begin{figure}[!bt]
  \centerline{
  \subfigure{\includegraphics[width=7cm]{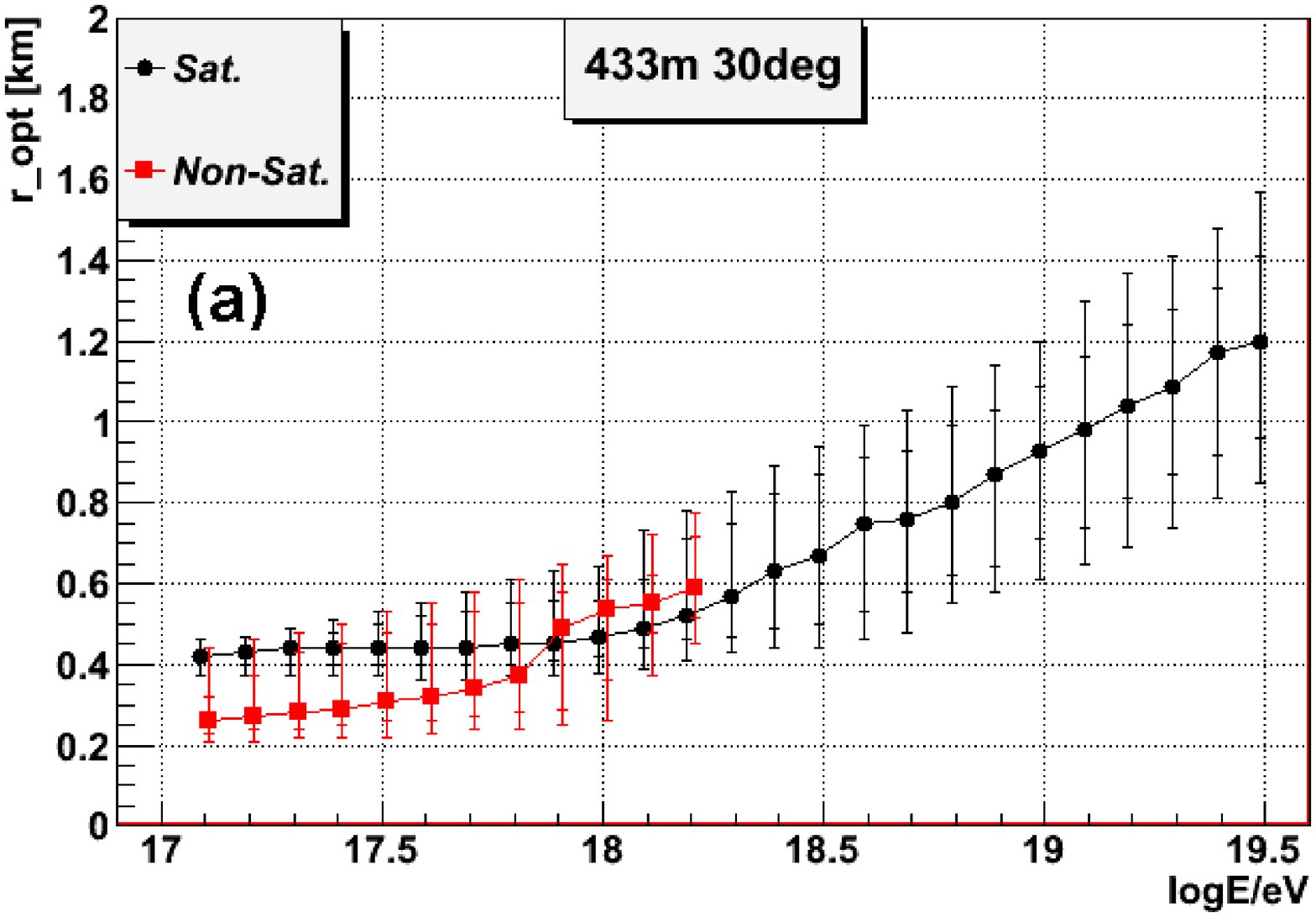}}
  \hfil
  \subfigure{\includegraphics[width=7cm]{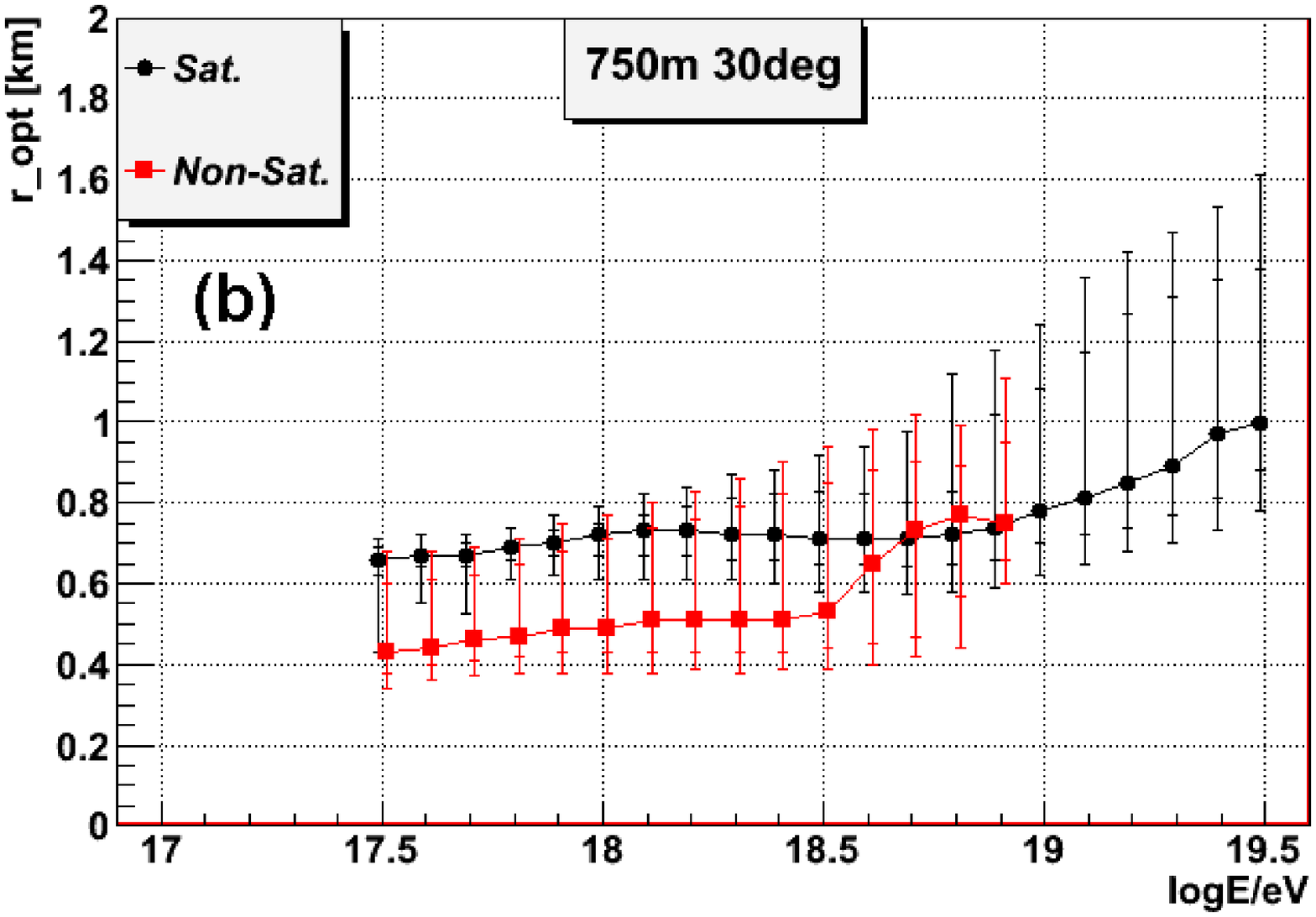}}
  }
  \centerline{
  \subfigure{\includegraphics[width=7cm]{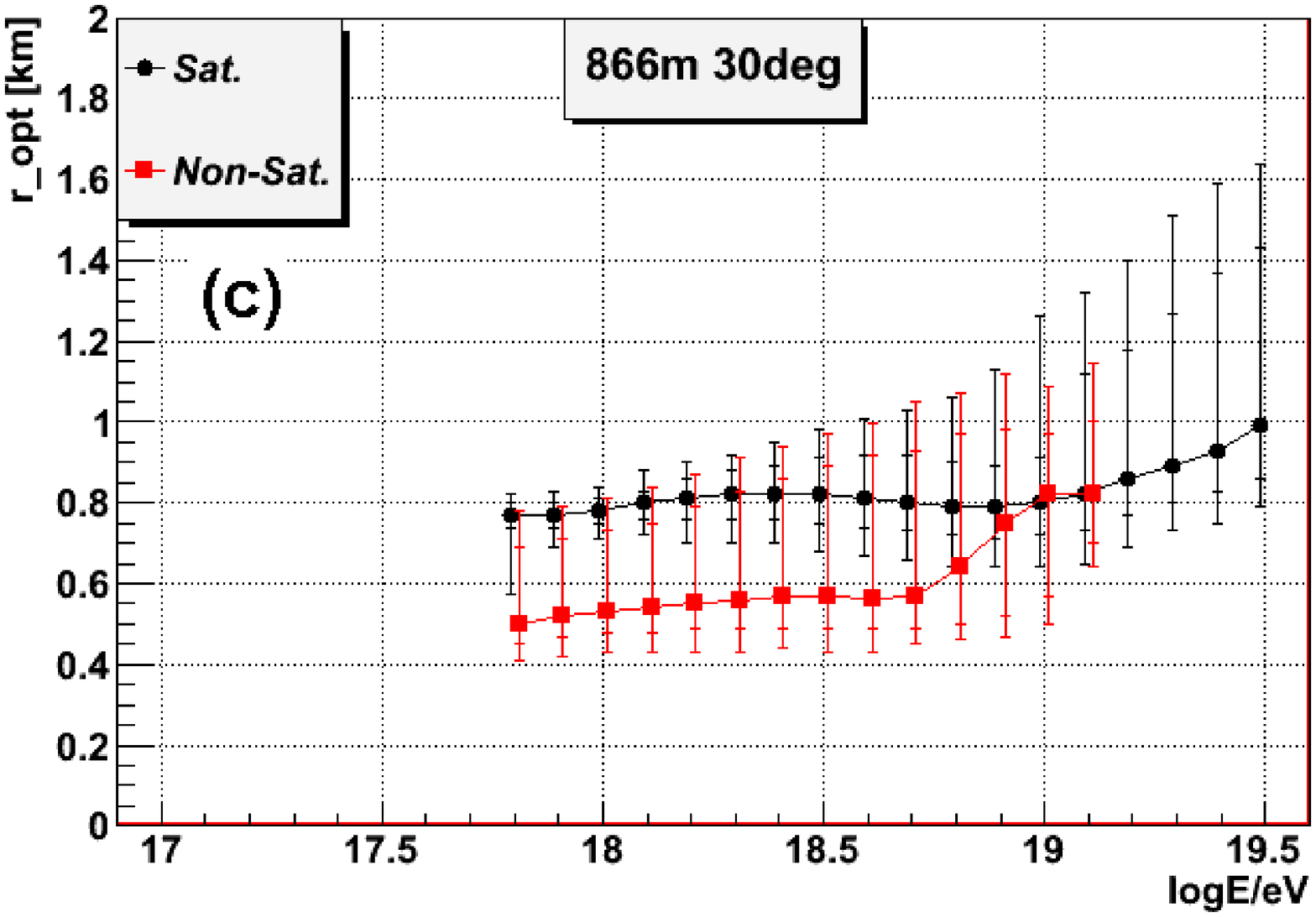}}
  \hfil
  \subfigure{\includegraphics[width=7cm]{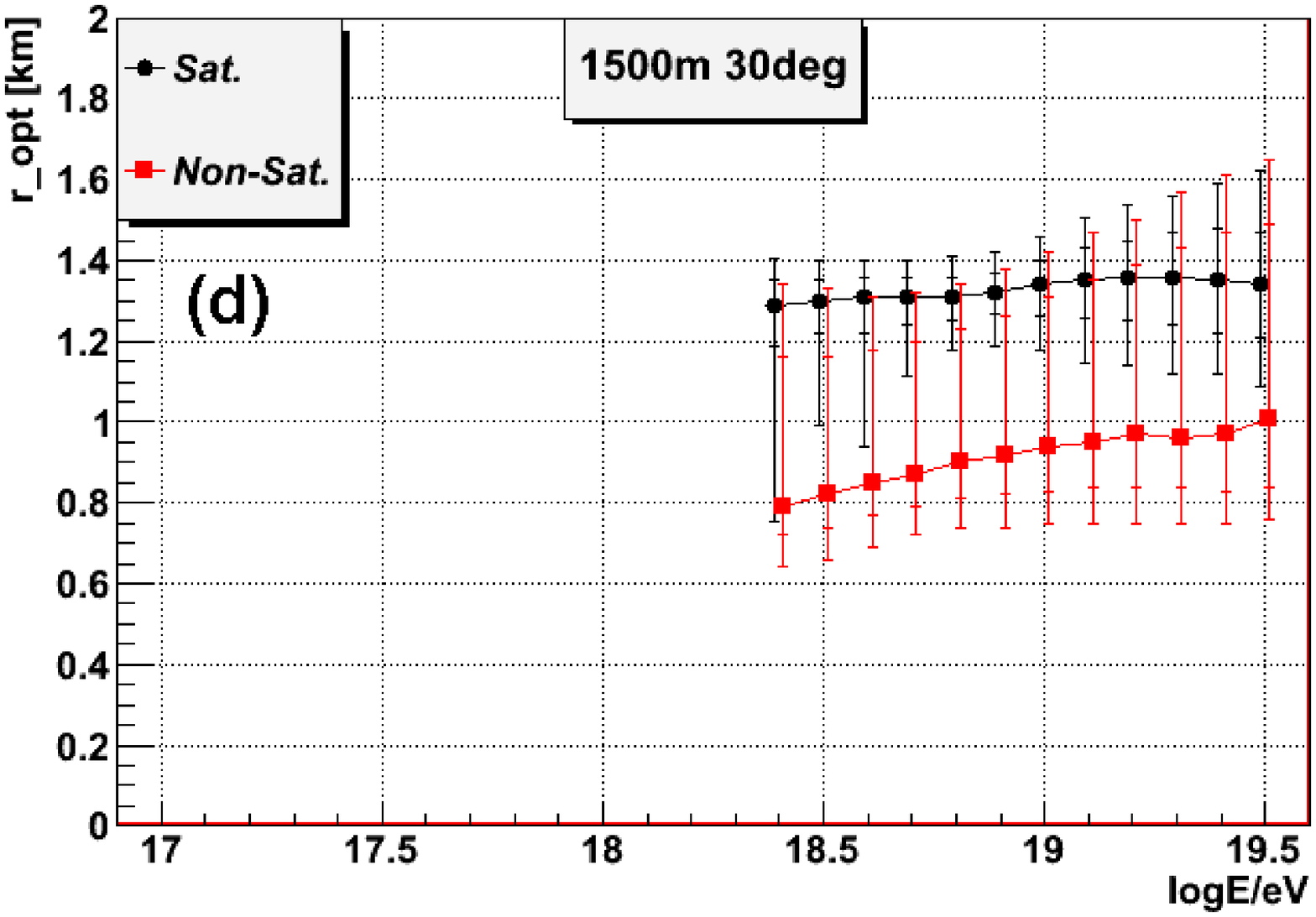}}
  }
  \caption{$r_{opt}$ vs. energy. Events with and without saturated detectors are shown separately. Zenith angle is $\theta=30^o$. The error bars represent the 68\% and 95\% C.L.}
  \label{fig:ropt_vs_energy_sat_non-sat}
\end{figure}

\begin{figure}[!bt]
  \centerline{
  \subfigure{\includegraphics[width=7cm]{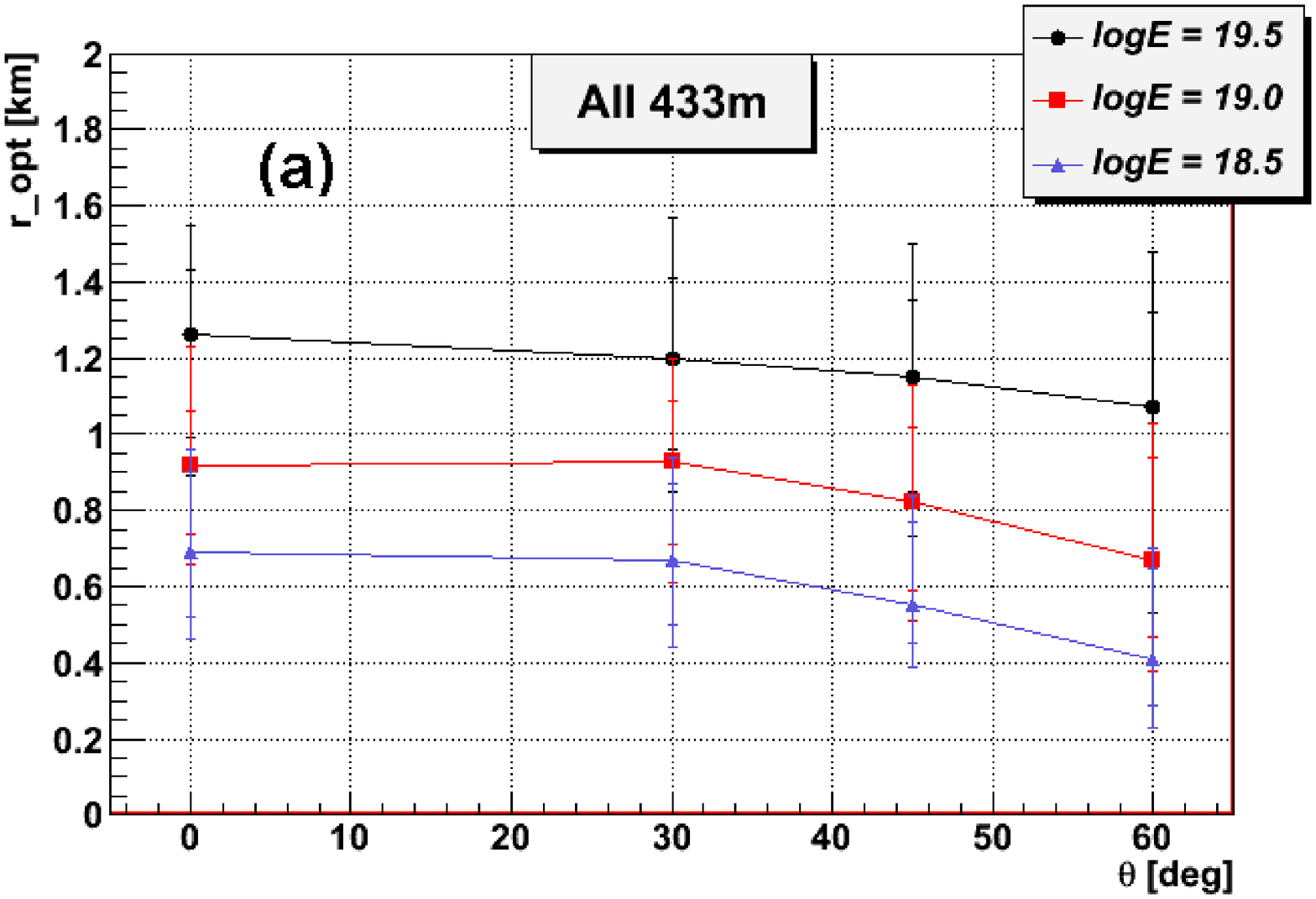}}
  \hfil
  \subfigure{\includegraphics[width=7cm]{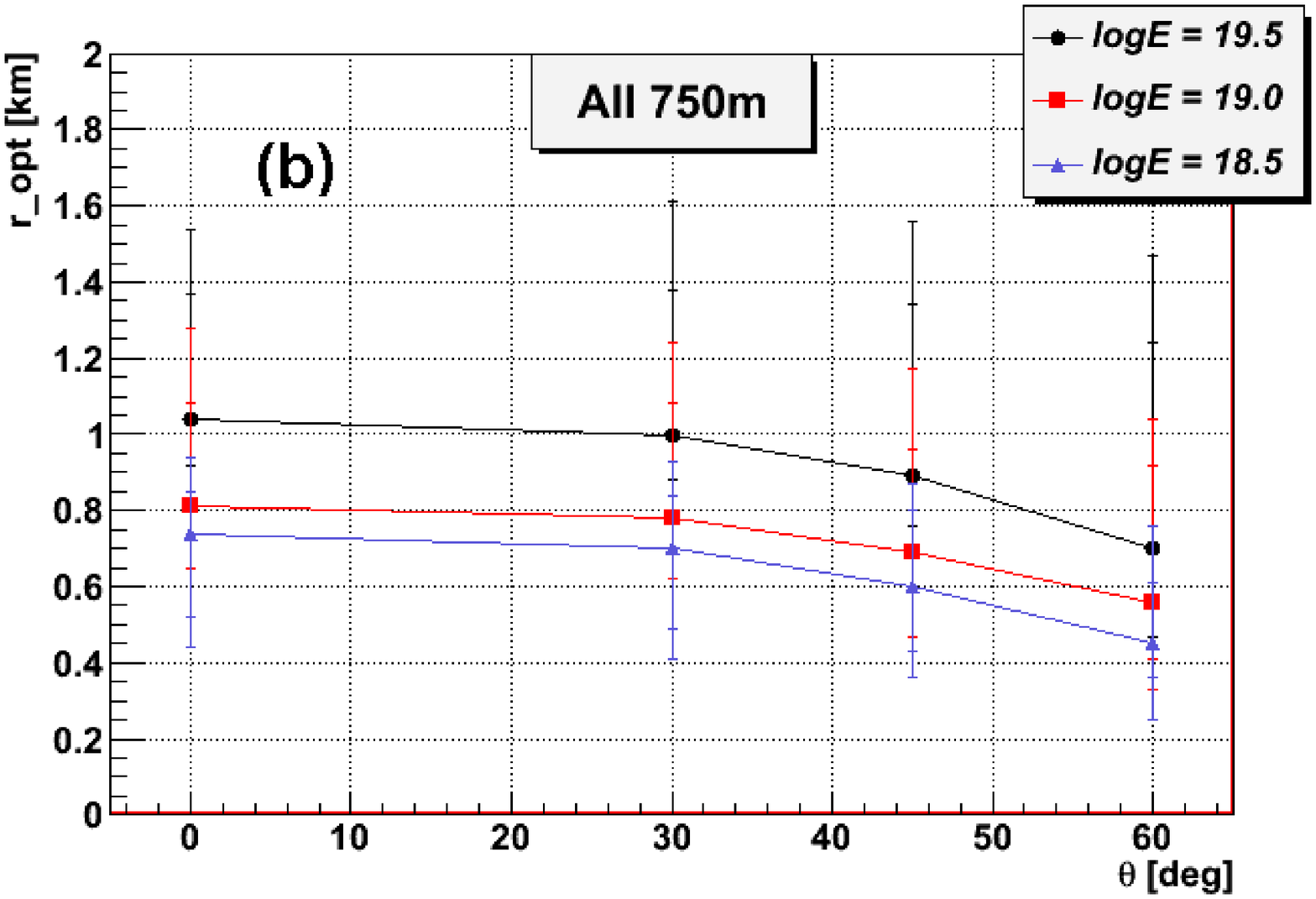}}
  }
  \centerline{
  \subfigure{\includegraphics[width=7cm]{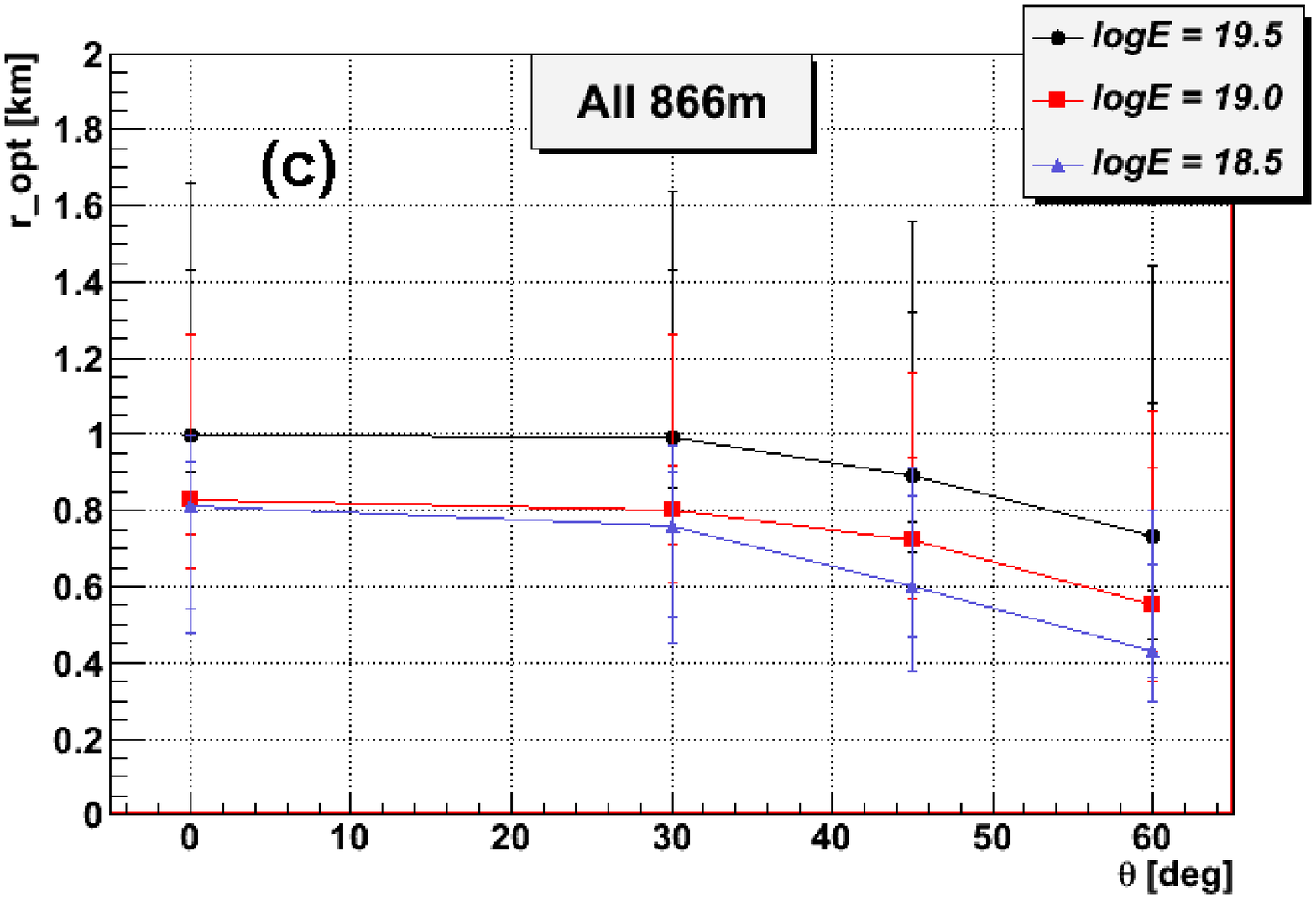}}
  \hfil
  \subfigure{\includegraphics[width=7cm]{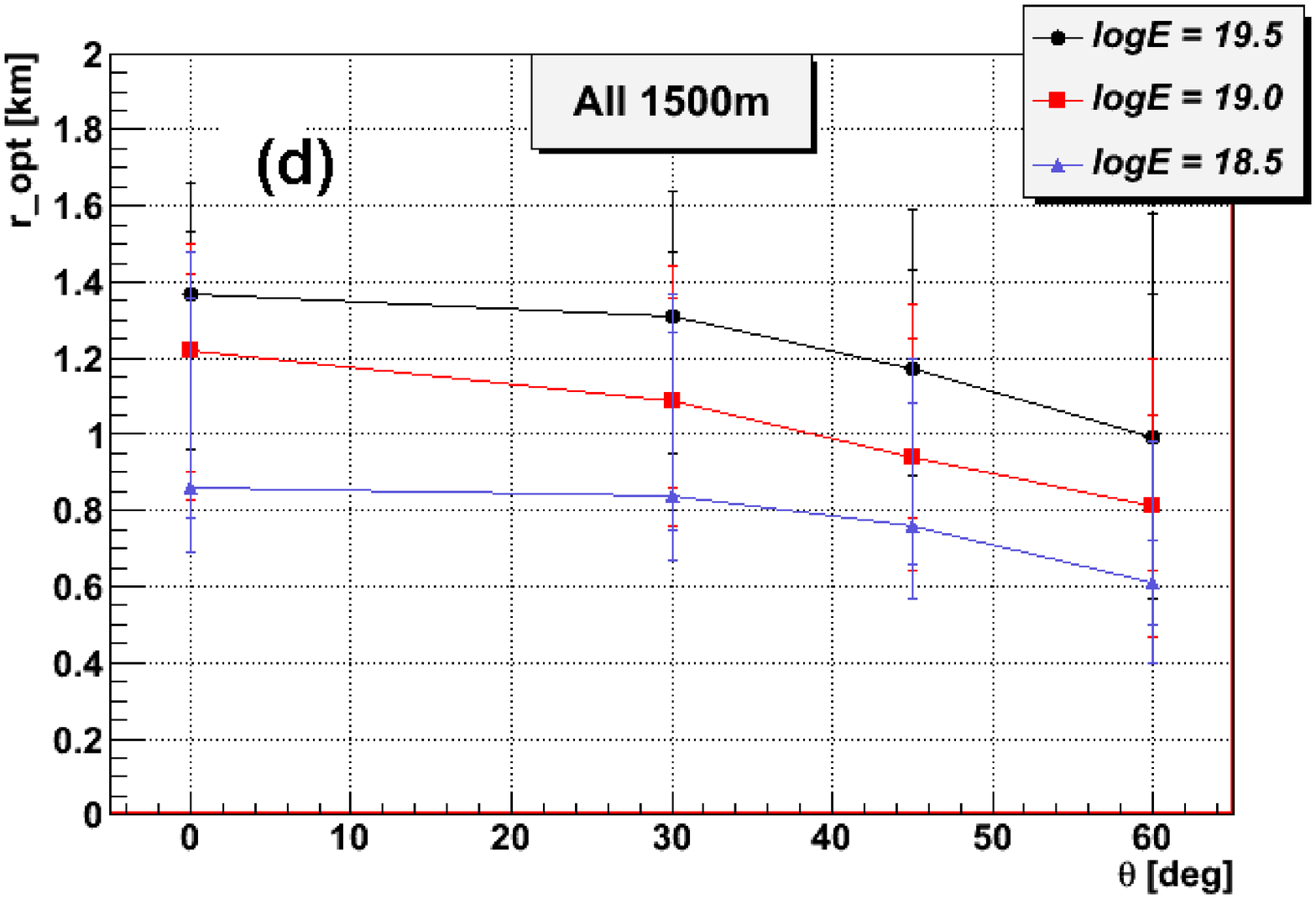}}
  }
  \caption{$r_{opt}$ vs. zenith angle for different array spacing and energies. The error bars represent the 68\% and 95\% C.L. Events with and without saturated detectors are both included.}
  \label{fig:ropt_vs_theta_all}
\end{figure}

\begin{figure}[!bt]
  \centerline{
  \subfigure{\includegraphics[width=7cm]{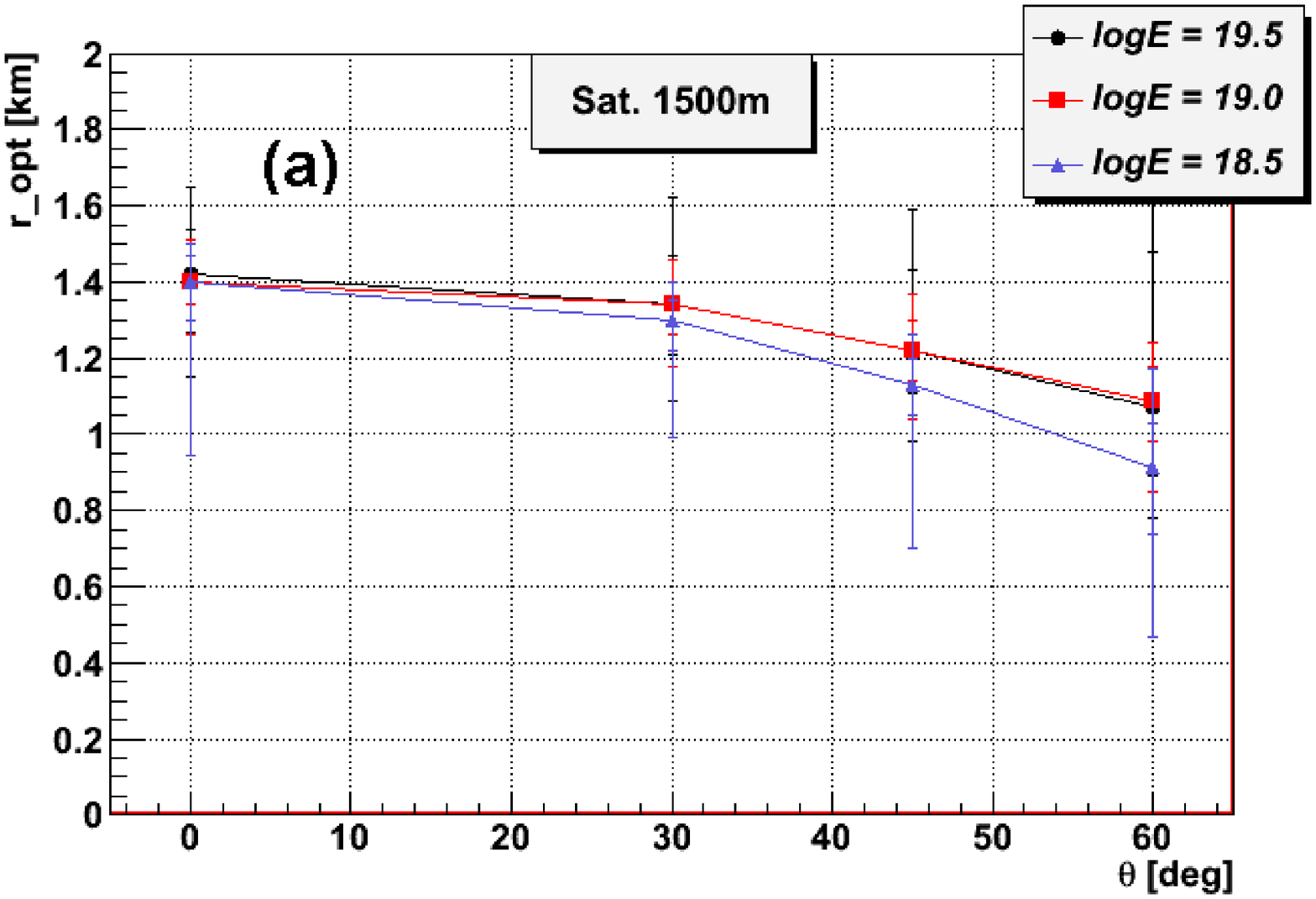}}
  \hfil
  \subfigure{\includegraphics[width=7cm]{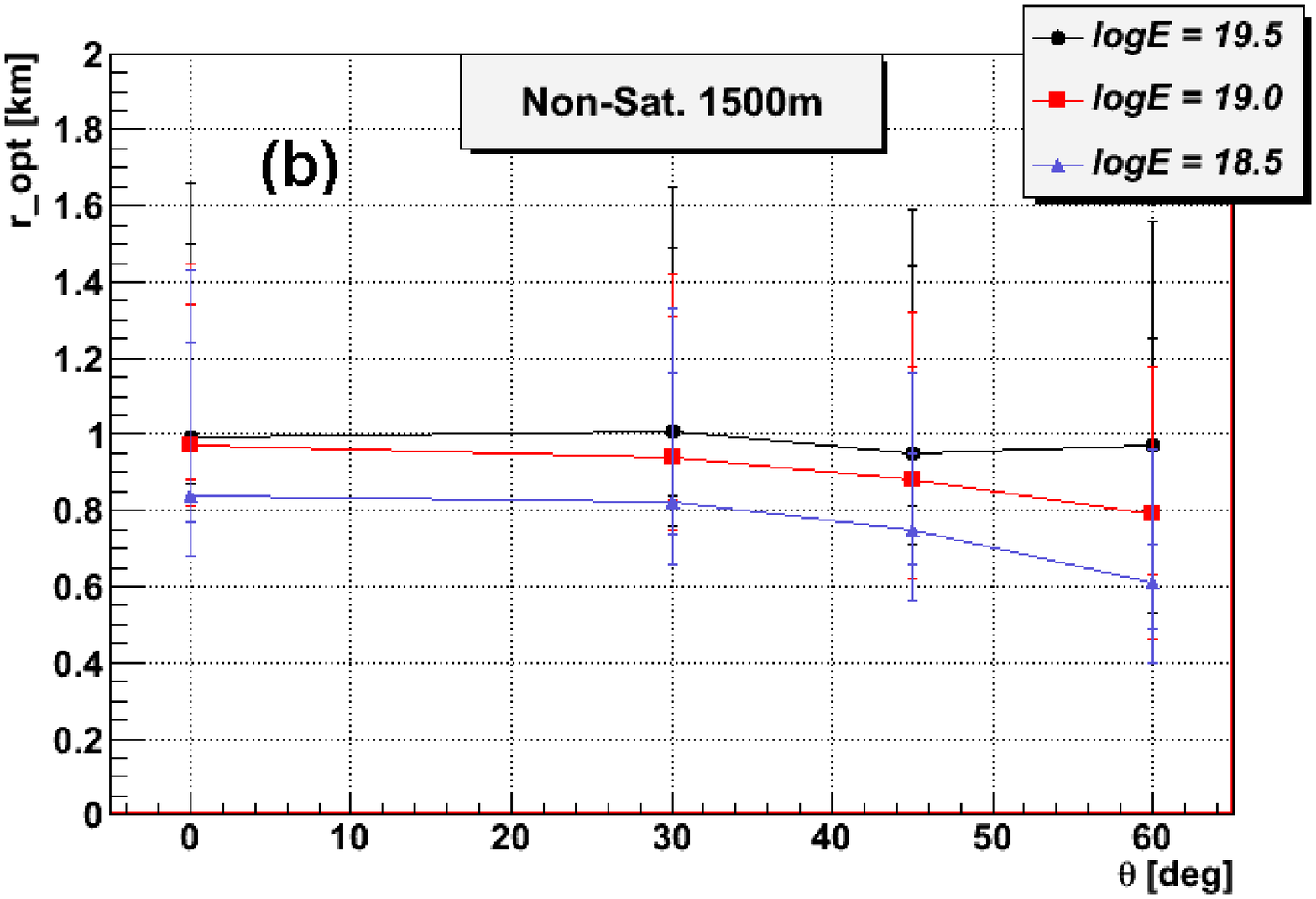}}
  }
  \caption{$r_{opt}$ vs. zenith angle for 1500 m separation array. The error bars represent the 68\% and 95\% C.L. (a): Events with saturated detectors. (b): Events without saturated saturated detectors..}
  \label{fig:ropt_vs_theta_sat_non-sat}
\end{figure}

Figure \ref{fig:ropt_vs_energy_all} shows the dependence of $r_{opt}$ with energy without discriminating whether events have saturated detectors (labeled as {\it All\/}). Showers are injected with zenith angles $\theta = 30^{o}$ and $60^{o}$.  It can be seen that $r_{opt}$ is a monotonous increasing function of energy due to the triggering of stations progressively further away from the core as energy increases. In Figure \ref{fig:ropt_vs_energy_sat_non-sat} the same behavior is shown separately for events with and without saturation.

Since events with and without saturated detectors may, and indeed do, behave in different ways, Figure \ref{fig:ropt_vs_energy_sat_non-sat} shows $r_{opt}$ results for both separately. All the previous array spacings are considered but only at one zenith angle, $\theta = 30^{o}$. $r_{opt}$ is an  increasing function of energy for both sets of events. It can be seen that $r_{opt}$ is larger for events with saturated detectors at lower energies ($r_{opt}^{sat} > r_{opt}^{non-sat}$) but that, at higher energies,  $r_{opt}^{non-sat}$ rapidly grows towards $r_{opt}^{sat}$. The transition energy region is narrow ($\Delta log E \sim 0.2$) and shifts upwards in energy as the array spacing grows. The symmetry of the triangular array with respect to a shower core located at the center of an elementary triangle (the least likely configuration to saturate at any given energy), manifest itself in ring-like arrangements of triggered stations. The appearance of a third ring of triggered stations is responsible for the rapid grow of $r_{opt}^{non-sat}$ over a limited energy interval as shower energy grows. 

Furthermore, low energy events with saturation have their cores very near the saturated stations. Therefore, the first triggered stations that do not saturate are clustered at the same distance from the core, which is roughly the array separation distance. 
Therefore, it is at the array separation distance that the dispersion among the several fits to the LDF is minimum. At higher energies, however, the next ring of the array enters into the set of triggered detectors of the event and, naturally, $r_{opt}$ increases. In Figure \ref{fig:ropt_vs_energy_sat_non-sat} it can be seen that $r_{opt}$ is almost constant and very near to the array separation at the lower energies, and that there is a threshold energy, which depends on the array separation, from which $r_{opt}$ increases steadily with energy. 

Figure \ref{fig:ropt_vs_theta_all}  shows the dependence of $r_{opt}$ with zenith angle for the same array spacings and three different input energies: $log(E/eV) = 18.5, 19.0$ and $19.5$. Both, events with and without saturated detectors are included. It can be seen that $r_{opt}$ is almost independent of zenith angle for $\theta \lesssim 30^{o}$ for any array spacing. However, as the zenith angle increases beyond $30^{o}$, $r_{opt}$ decreases with $\theta$, independently of array spacing and energy. The same effect is observed in both sets of events, those with saturation (Figure \ref{fig:ropt_vs_theta_sat_non-sat}.a) and without it (Figure \ref{fig:ropt_vs_theta_sat_non-sat}.b). This result comes from the fact that, for inclined showers, the array spacing projected onto the shower front shrinks with zenith angle and $r_{opt}$ naturally follows this behavior.

From the previous results, it is clear that $r_{opt}$ is in general a function of energy and zenith angle for inclined showers. Furthermore, in Figures \ref{fig:ropt_vs_energy_all}, \ref{fig:ropt_vs_energy_sat_non-sat},  \ref{fig:ropt_vs_theta_all} and \ref{fig:ropt_vs_theta_sat_non-sat}, the error bars indicate the $68$\% and $95$\% confidence levels (CL) and the central points correspond to the median value of $r_{opt}$. It can be seen that, in all cases, even if the behavior of the median curves is rather smooth, the CL are large and, therefore, considerable fluctuations are expected. Additionally, due to the large relative fluctuations of the signals from detectors located at large distances from the shower axis, the error distributions are skewed in general towards larger values from the median of $r_{opt}$. These points argues strongly in favor of an $r_{opt}$ determined specifically for each shower since, using a fixed characteristic value, $r_{0}$, could compromise the estimation of primary energy. This possibility is analyzed in the following section.

As it was mentioned in the introduction, although similar in character, the work in reference \cite{Newton} is rather different in algorithmic approach and scope. Therefore, a comparison between results in both works is not straightforward. Nevertheless, 
Figures 5 in \cite{Newton} can be used to some extent to crosscheck our results. 
Figure 5 bottom-right in \cite{Newton} shows their $r_{opt}$ as a function of energy. 
Despite the fact that there are indications of border effects at low energies in their calculation and that different zenith angle events are binned together, the results are similar to those in our Figure \ref{fig:ropt_vs_energy_sat_non-sat}.d. Figure \ref{fig:ropt_vs_energy_sat_non-sat} shows $r_{opt}$ for events with and without saturated detectors in the energy interval between $\sim 10$ and $30$ EeV, for 433 (a), 750 (b), 866 (c) and $1500$ m (d) spacing. It can be seen that, at 433, 750 and 866 m spacing $r_{opt}$ is more or less independent of energy at lower energies but eventually increases steadily above a certain energy. This effect is also expected at a separation of 1500 m for energies beyond those presently plotted in Figure \ref{fig:ropt_vs_energy_sat_non-sat}.d. Reference \cite{Newton}, on the other hand, shows results only for arrays at 1500 m separation, where the same trend seems to be suggested for events with saturated detectors (see Figure 5 bottom-right of \cite{Newton}). Remarkably, although their analysis extends up to 100 EeV, the same trend is not seen for events without saturation. The latter, however, may be due to the fact that in reference \cite{Newton} showers with all zenith angles are mixed together which, at high energies, implies that their sample must be highly biased to very inclined events (otherwise they would present saturation), masking the effect. In fact, it can be seen from our Figure \ref{fig:ropt_vs_theta_sat_non-sat}.b that, for events without saturation, $r_{opt}$ does
decrease at any energy for larger zenith angles.

Again, in Figure 5 bottom-left of \cite{Newton}, and despite the fact that the authors claim only a slight dependence of $r_{opt}$ with zenith angle, we obtain a very similar result for events with saturation
in Figure \ref{fig:ropt_vs_theta_sat_non-sat}.a with $r_{opt}$ decreasing markedly with increasing 
zenith angle. There is no agreement, however, for events without saturation, where they obtain an
$r_{opt}$ that increases with zenith angle, while our results (see, Figure \ref{fig:ropt_vs_theta_sat_non-sat}.b)
shows an $r_{opt}$ that at low energies decreases as a function of zenith angle, but tends
to a constant value as the energy increases. Part of the difference between both results may be
due to the fact that in \cite{Newton} energies randomly selected from a flat spectrum have been
binned together. The latter, however, cannot account for their unexpected raise with zenith angle.

\section{Influence of $r_{opt}$ on reconstructed energy}

In this Section we analyze the effect of adopting a fix characteristic distance, $r_0$, instead of a shower-specific value, $r_{opt}$, for the determination of shower energy and energy spectrum. We simulate a detector similar to AGASA (see Section \ref{Algorithm}), i.e., a separation of 1 km between stations and use eq. \ref{AgasaLDF} and  eq. \ref{Scintillators_LDFFit} in order to generate signals and fit the ``observed'' LDF respectively. For each event $r_{opt}$ is estimated using the procedure explained in Section \ref{Algorithm} while eq. \ref{AgasaLDF} is used in order to estimate the energy for both $r_{0}=600$ m, as AGASA did, and $r_{opt}$. 

Two different input spectra are used. A spectrum with one thousand events per energy bin ($\Delta \log(E)=0.1$) from $10^{17.8}$ to $10^{20.7}\; eV$, is used in order to study the functional form of the energy error distributions and the energy reconstruction bias (Sections 4.1 and 4.2). The energy reconstruction of events with saturated detectors is also analyzed. Second, in Section 4.3, a more structured spectrum extending from $10^{17.7}$ to $10^{20.5} \; eV$, which possesses an ankle, a GZK-cut-off, and is exposure-limited at low energy, is used to assess the impact of both techniques in a more realistic situation. For every event, the angular distribution is extracted randomly from an isotropic distribution with a maximum zenith angle of $45^{o}$, as in the case of the AGASA experiment. The azimuthal angles are selected from a uniform distribution between $0^{o}$ and $360^{o}$ and the core location is randomly located inside an elementary cell.

It must be noted that the results of this Section do not directly apply to the spectrum inferred from  surface arrays that relay on the use of hybrid events for the energy calibration. The main reason is that the error in core location for hybrid events is much smaller than for pure surface events. In the case of Auger, for example, the error in hybrid core determination is only around $35$ m \cite{AugerHybPerf07} while for the majority of pure SD events it is estimated to be around $100$ m \cite{AugerUncRec05}. Therefore, $r_{opt}$ for hybrid events is very much constrained. In addition, hybrid experiments do not directly relate the signal measured at $r_{opt}$ with the primary energy. Instead, they use a calibration with the energy obtained by the fluorescence technique. Finally, the most important uncertainties in the primary energy determination in hybrid experiments come from the fluorescence uncertainties not from the parameter size determination as will be discussed later in detail.

\subsection{Error distribution function in energy reconstruction}\label{SectionGauss}

We calculate the distribution function of the errors in reconstructed energy, i.e. the difference between the reconstructed and the real energy, as a function of injected energy for both techniques, $r_0$ and $r_{opt}$. 

\begin{figure}[!bt]
  \centerline{
  \subfigure{\includegraphics[width=7cm]{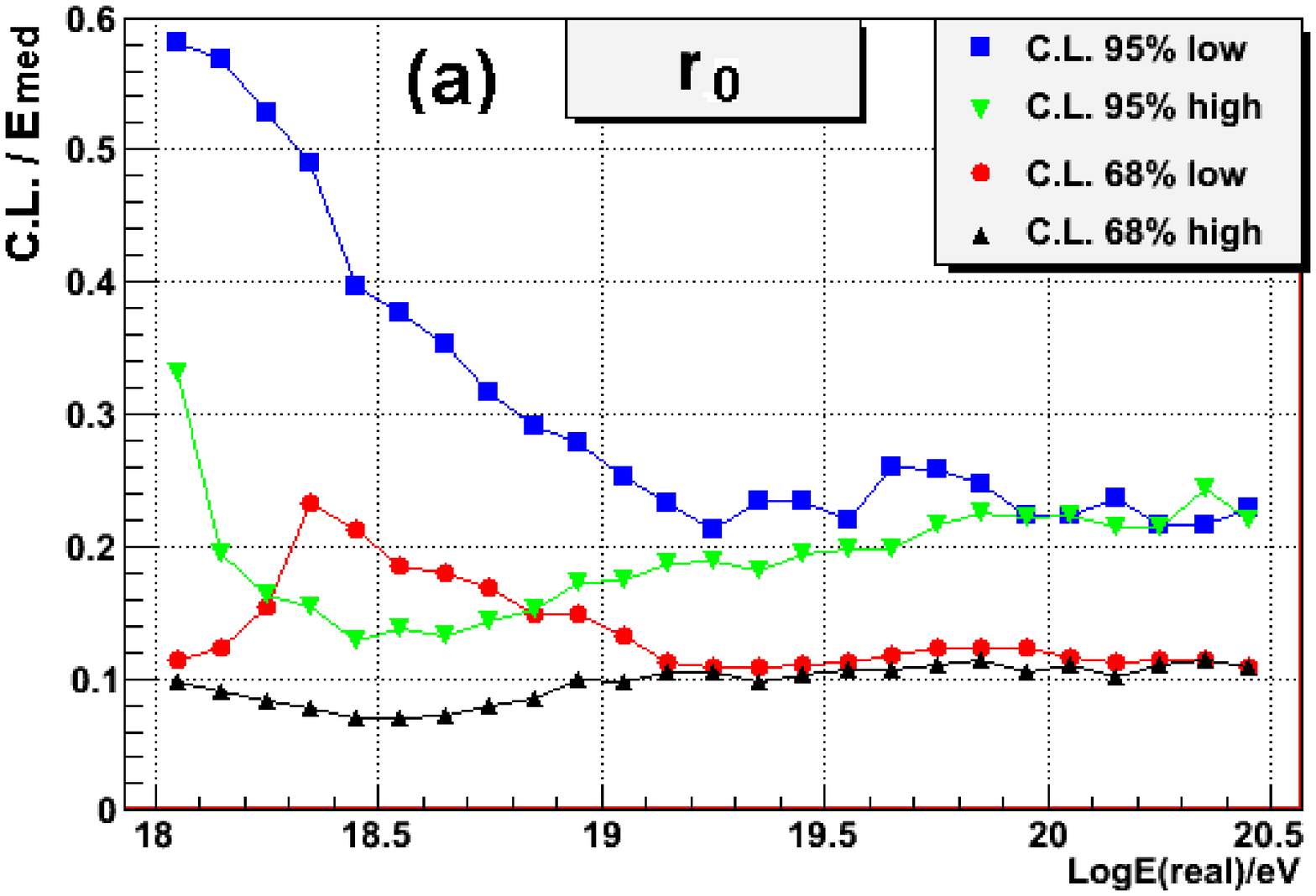}}
  \hfil
  \subfigure{\includegraphics[width=7cm]{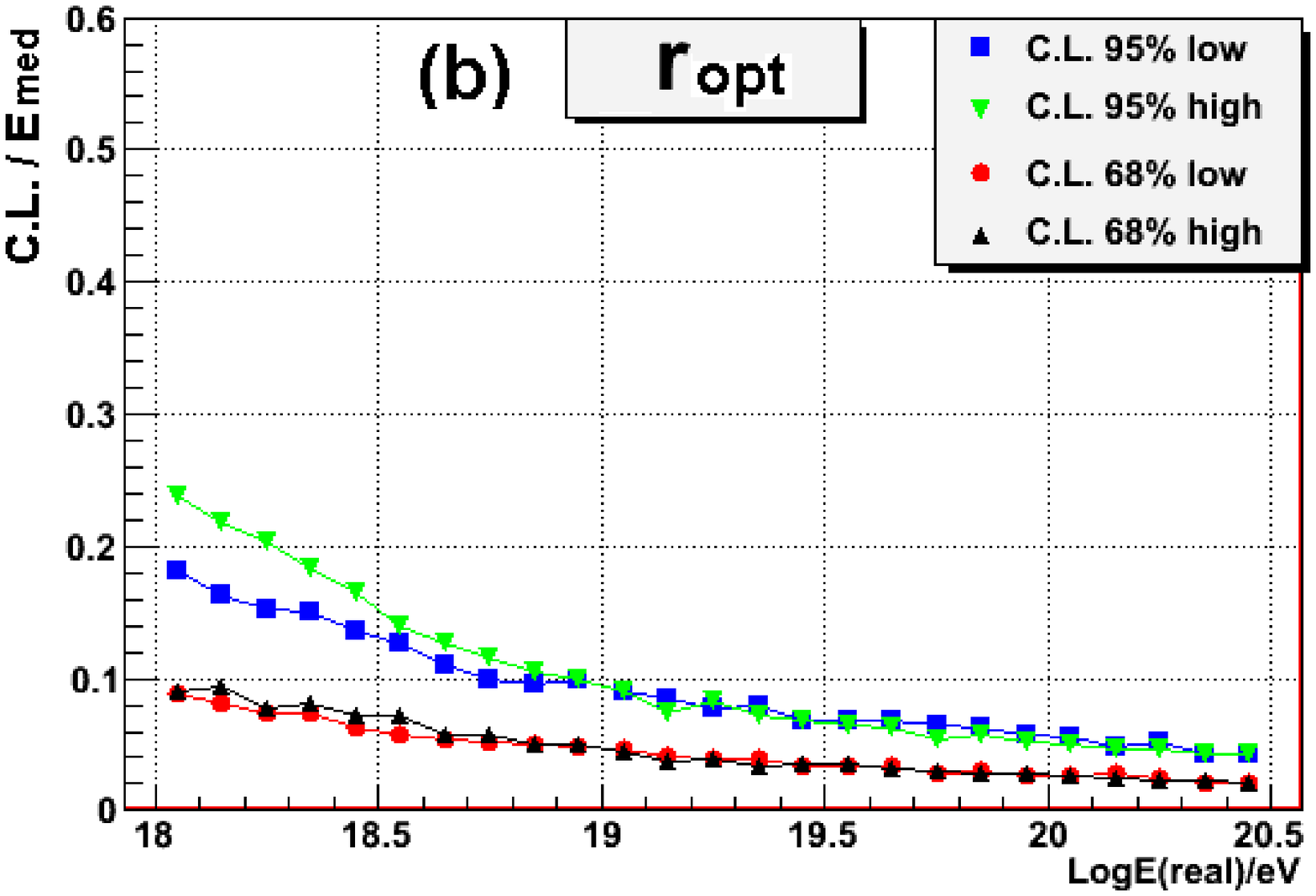}}
  }
  \caption{68\% and 95\% CL over the median, from both its lower and higher energy sides, for the energy error distribution functions determined using either $r_0$ (a) or $r_{opt}$ (b) methods. See text for more details.}
  \label{fig:rec_errors}
\end{figure}

Figure \ref{fig:rec_errors} shows, for both $r_0$ (a) and $r_{opt}$ (b), the 68\% and 95\% CL for the right and left sides with respect to the median of the energy error distribution. It can be clearly seen that the error distribution functions originated by using $r_{opt}$ are much more compact and symmetrical than the corresponding distributions for $r_{0}$. The effect is more notable for lower energies where the distribution function for characteristic distance determination is particularly wide and skewed. Although these figures are drawn for the $1000$ m separation array, the results apply qualitatively for any of the other spacings considered in previous sections.

Arguably, it is desirable that the errors in energy reconstruction have a Gaussian distribution. Gaussian errors, for example, are easier to handle and understand when applying deconvolution techniques in the spectrum determination while assuring that there are no asymmetries or long tails, which further reduces the danger of border effects and biases associated with a rapidly changing spectral index. Again, it can be seen from Figure \ref{fig:rec_errors}.b that the $r_{opt}$ method produce, at any given input energy very approximately normal distributions.

\subsection{Bias in the reconstructed energies}

\begin{figure}[!bt]
  \centerline{
  \subfigure{\includegraphics[width=7cm]{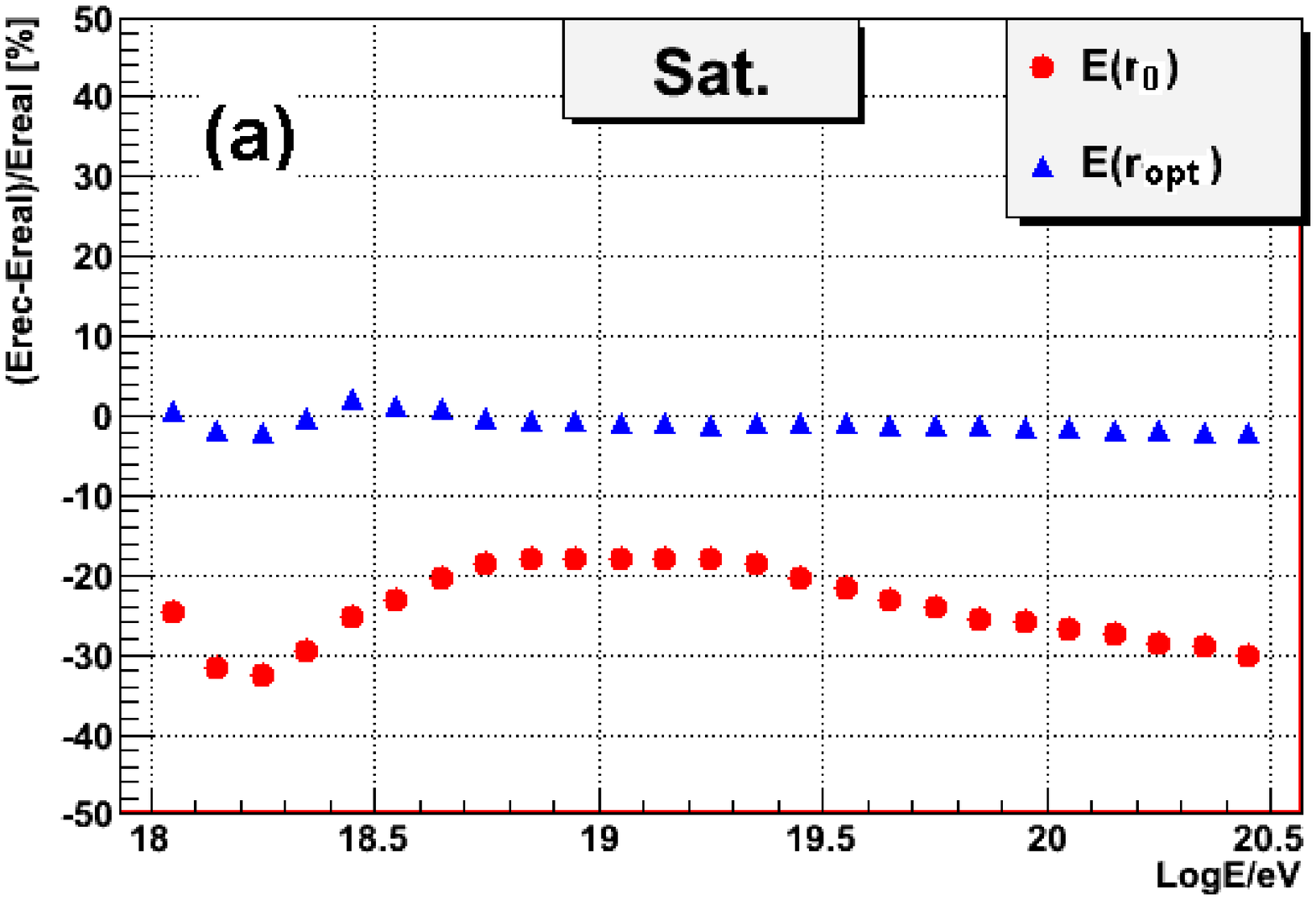}}
  \hfil
  \subfigure{\includegraphics[width=7cm]{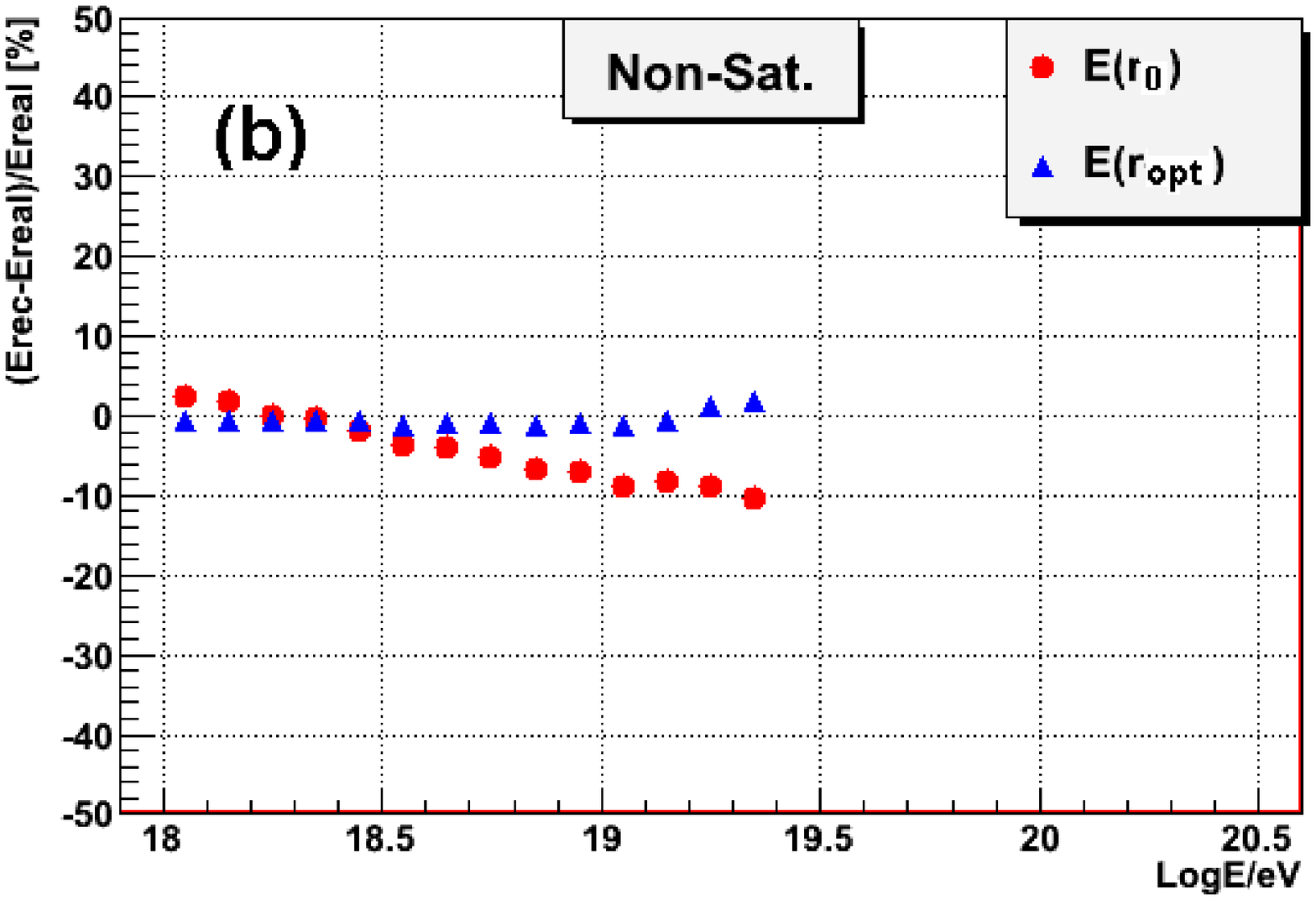}}
  }
  \caption{Bias in the reconstruction methods for events with (a) and without (b) saturated detectors.}
  \label{fig:Bias}
\end{figure}

Figure \ref{fig:Bias} shows the relative reconstruction error as a function of the injected energy for both reconstruction techniques. In the case of events without saturated detectors (Figure \ref{fig:Bias}.b), there is no appreciable bias using $r_{opt}$, while using $r_{0}$ there is an energy dependent bias which, at larger energies, can reach $\sim 10$\%. The difference is much more significant in the case of events with saturated detectors (Figure \ref{fig:Bias}.a): the $r_{opt}$ approach produces almost negligible bias in the whole energy range while the reconstructed energy is largely underestimated using $r_{0}$. The 
later underlines the fact that $r_{opt}$ is very different for both populations of events. Analogous results 
are obtained for any array grid size.

\begin{figure}[!bt]
  \centerline{
  \subfigure{\includegraphics[width=7cm]{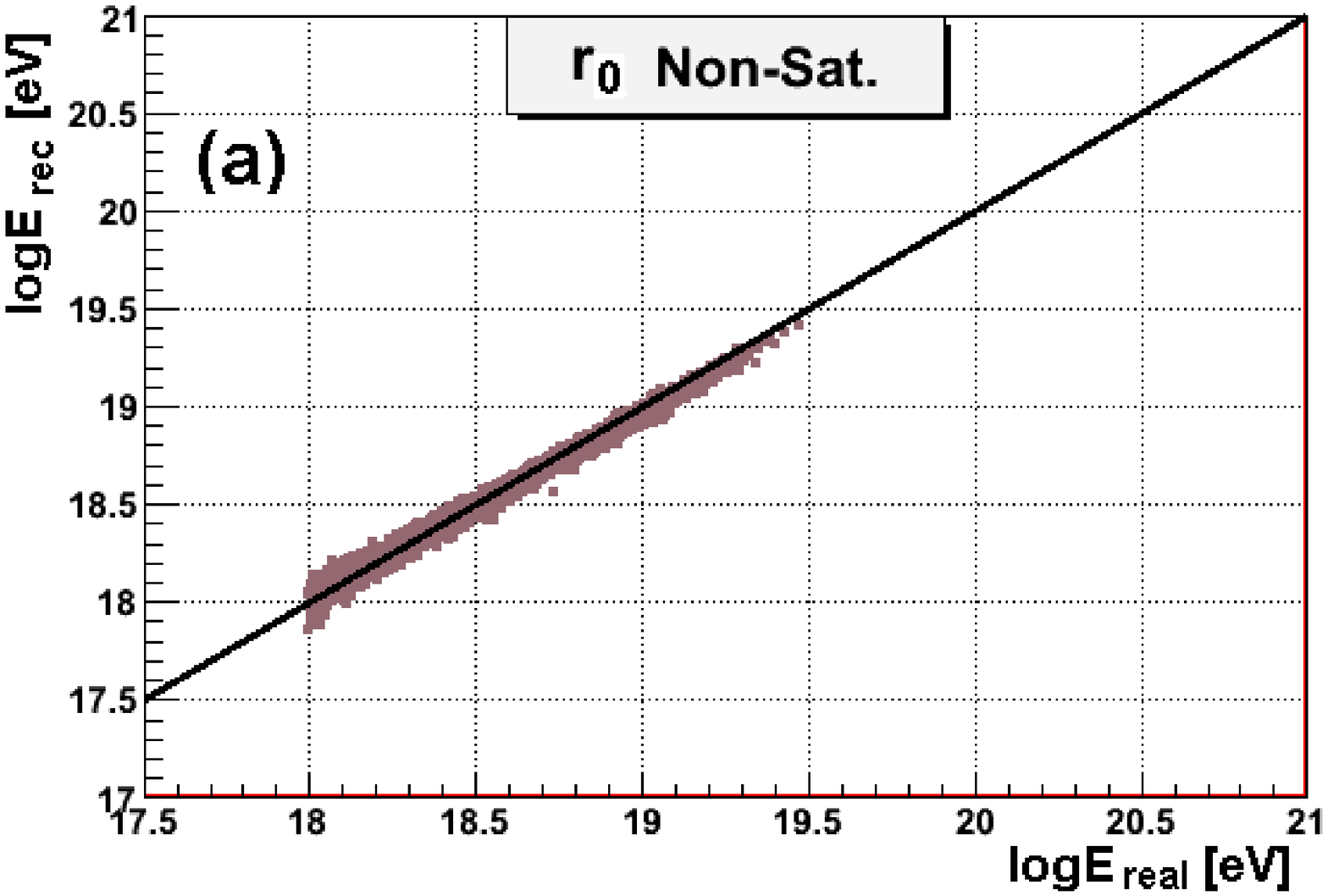}}
  \hfil
  \subfigure{\includegraphics[width=7cm]{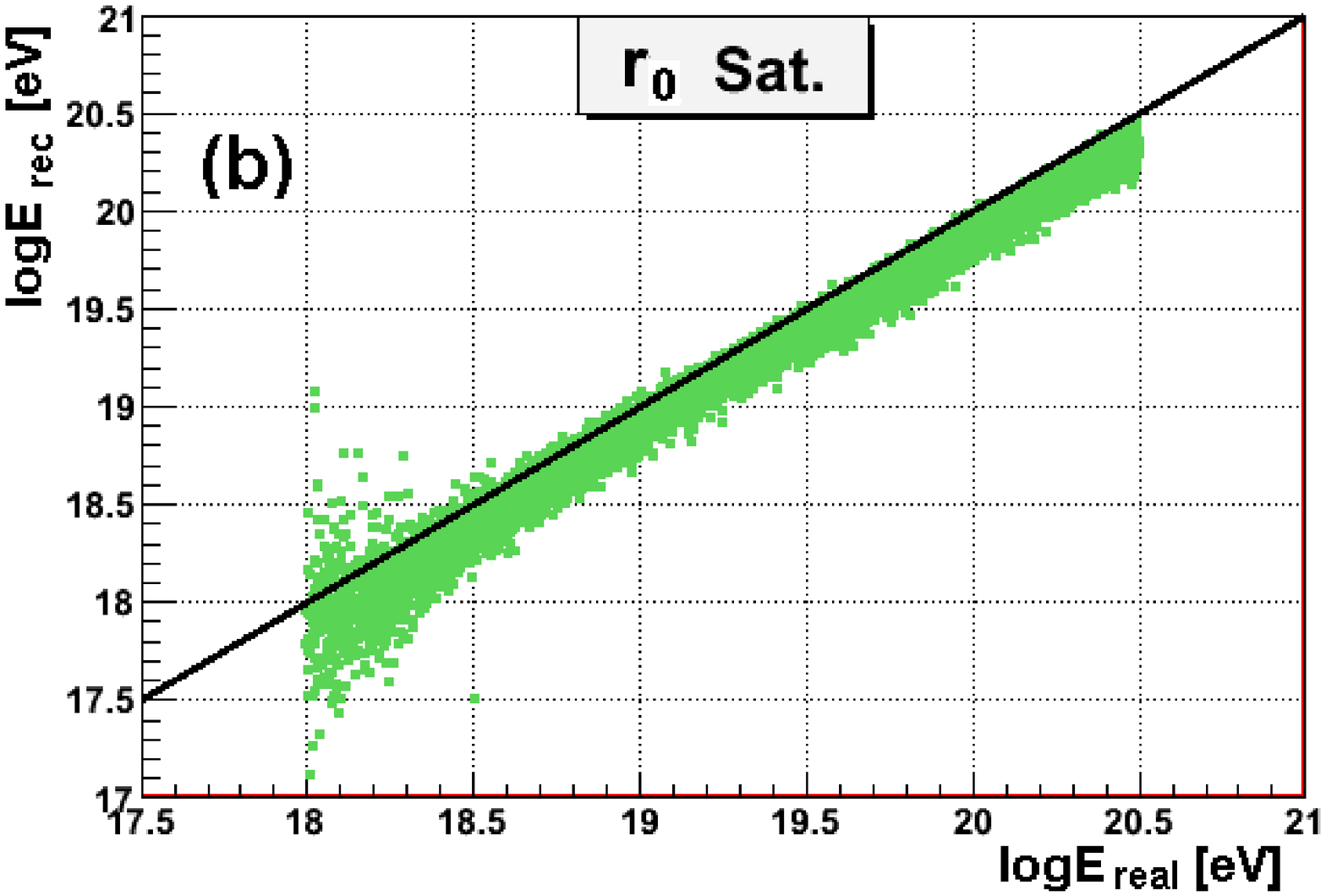}}
  }
  \centerline{
  \subfigure{\includegraphics[width=7cm]{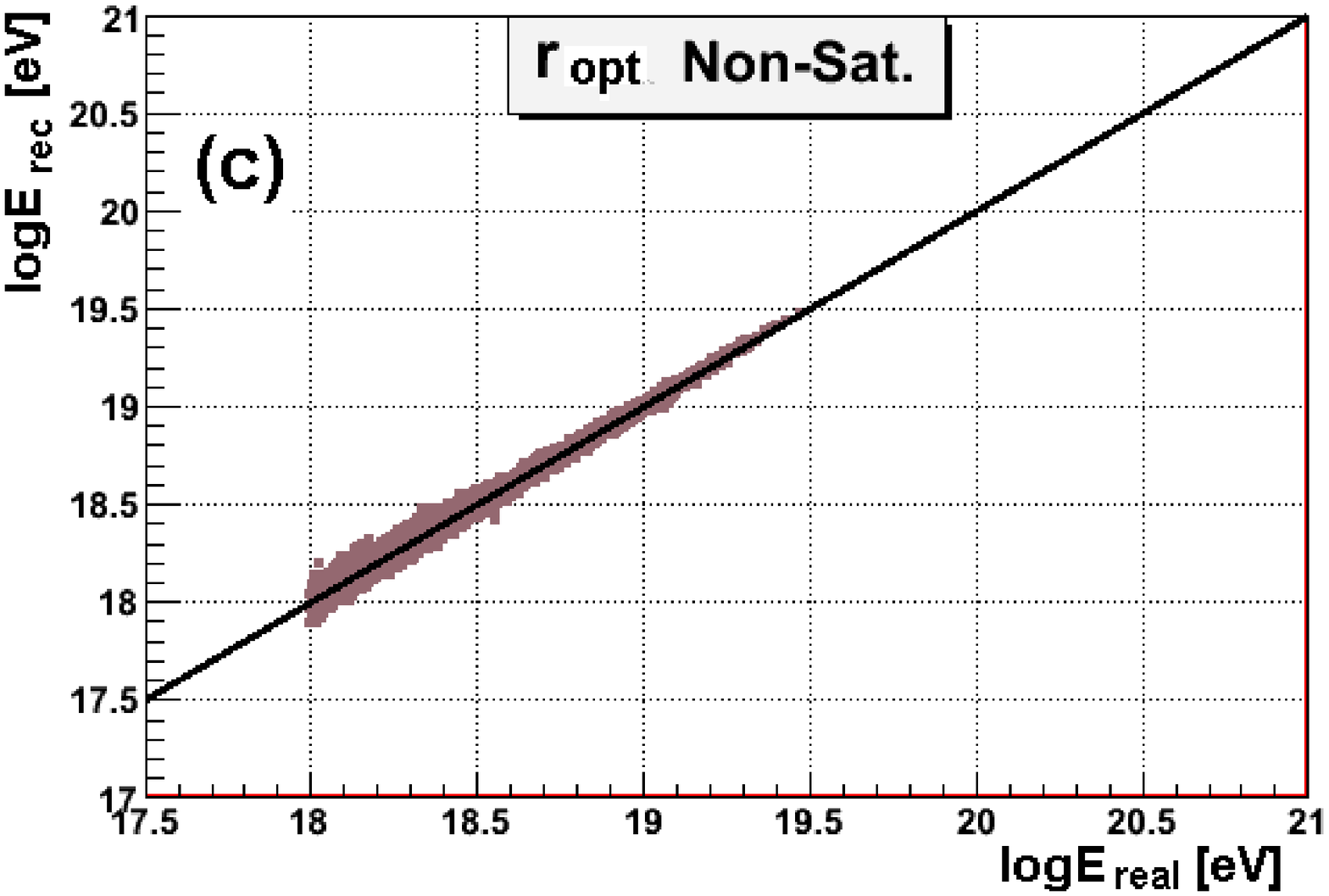}}
  \hfil
  \subfigure{\includegraphics[width=7cm]{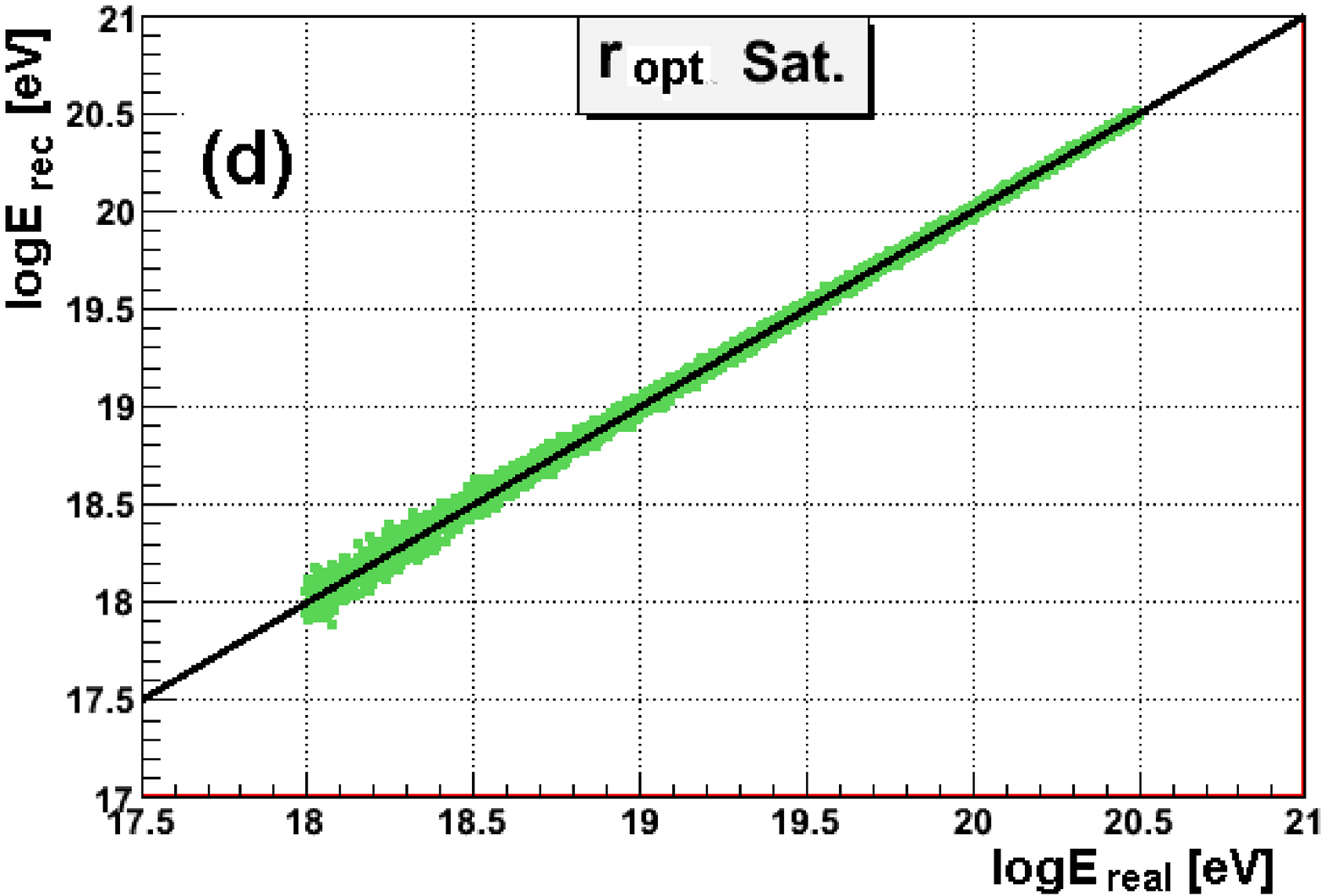}}
  }
  \caption{Reconstructed energy vs. real energy. Top: using $S(r_0)$. Down: using $S(r_{opt})$. Left: events without saturation. Rigth: events with saturation.}
  \label{fig:logErec_vs_logEreal}
\end{figure}

The energy reconstructions of events with and without saturated detectors are shown separately for both techniques in the scatter plots of Figure \ref{fig:logErec_vs_logEreal}. As commented before, better reconstruction is achieved using $r_{opt}$ for both classes of events, while a more significant difference appears for events with saturated detectors. 

In fact, a main advantage of our proposed method is the treatment of events with saturated detectors, which is shown for $r_{0}$ and $r_{opt}$ in Figures \ref{fig:logErec_vs_logEreal}.b and \ref{fig:logErec_vs_logEreal}.d respectively. Using a fix characteristic value $r_0$, events with saturation are poorly reconstructed, specially at lower energies. Essentially, the main problem is that these events have very few triggered stations and almost at the same distance from core. Consequently, their reconstruction accuracy is quite poor -- this is particularly true for the fit to the LDF. In practice, using the $r_{0}$ approach, these events probably would not pass the usual quality cuts and would be discarded, or would be reconstructed with a specific procedure. Nevertheless, the $r_{opt}$ approach makes it possible to infer without almost any bias the energy of all the events, with an accuracy comparable to that attained for events without saturation. 

\begin{figure}[!bt]
  \centerline{
  \subfigure{\includegraphics[width=7cm]{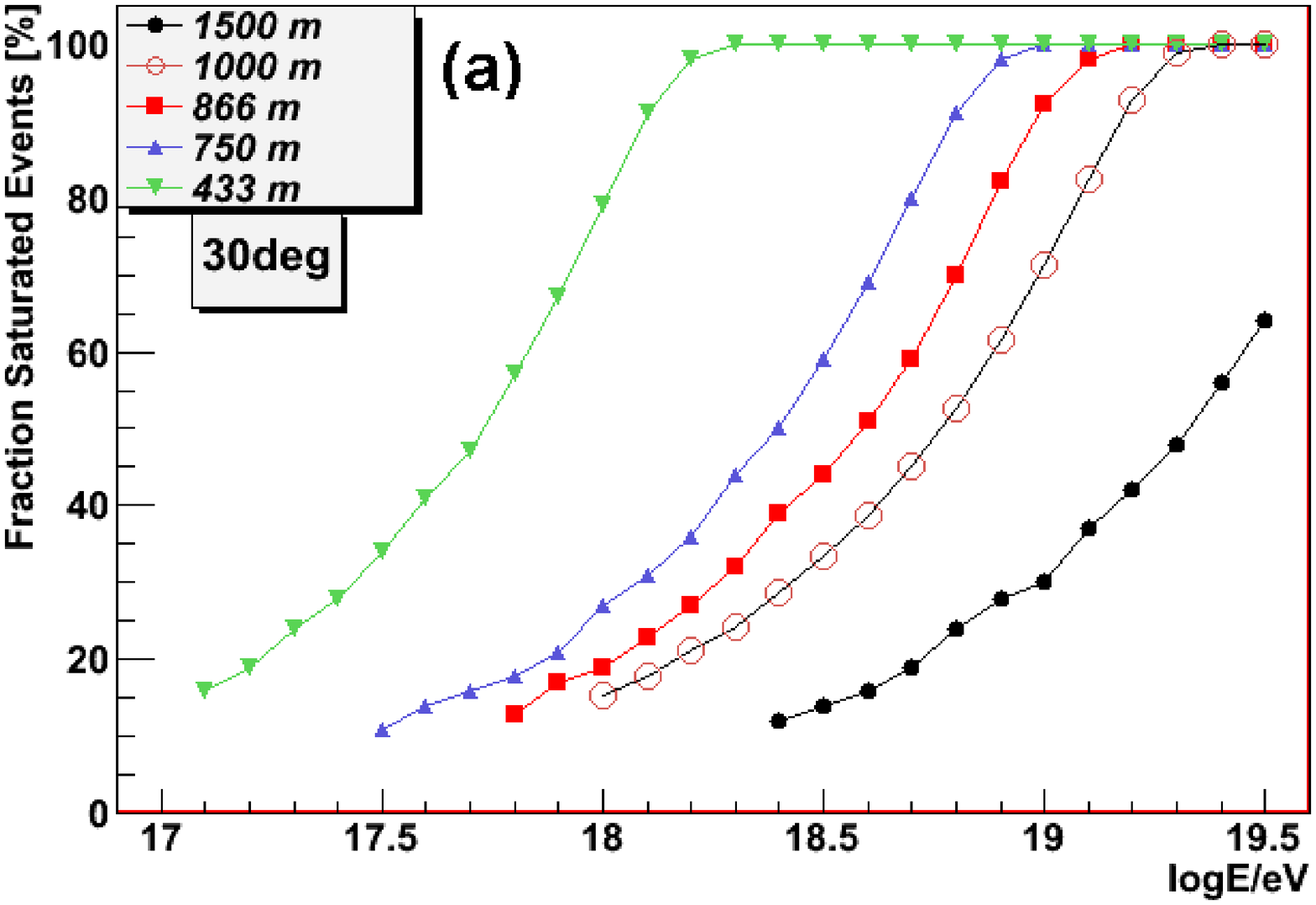}}
  \hfil
  \subfigure{\includegraphics[width=7cm]{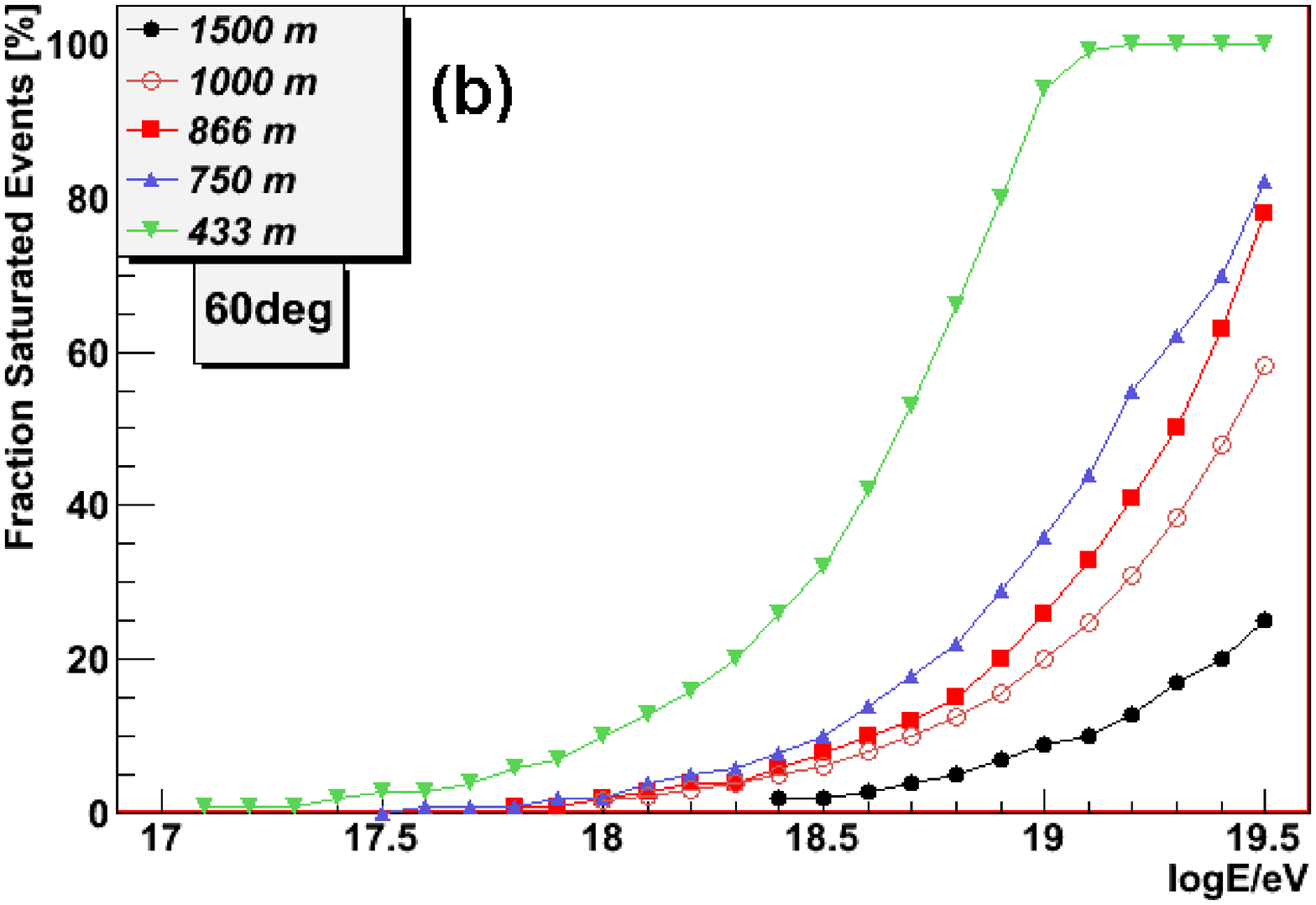}}
  }
  \caption{Fraction of events with saturated detectors as a function of energy for the different array spacings considered. (a) $\theta=30^o$. (b) $\theta=60^o$.}
  \label{fig:fracsat}
\end{figure}

The advantage of a homogeneous treatment for all classes of events is further stressed by the fact that events with saturated detectors are in general dominant for most of the operational range of an experimental array, regardless of the detector separation (see Figure \ref{fig:fracsat}). For example, considering the 1 km separation array used for these studies, the number of triggered detectors in an event varies from 5 to 60 for showers from $10^{17.5}$ to $10^{19.5}$ eV. Considering an incoming event of $E \sim 10^{19.5}$ eV and a zenith angle of $\theta \sim 30$ degrees, any detector located at $< 550$ m from the shower axis would very likely be saturated. Thus, independently of the position of the core inside the elementary cell, almost 100\% of the events have at least one saturated detector. At still higher energies, even 2 or
3 detectors could be saturated. Furthermore, for the same spacing, 50\% of the events will have at least one
saturated detector above $E \sim 6$ EeV (see Figure \ref{fig:fracsat}).

\subsection{Reconstruction of a rapidly changing spectrum}

In the previous section we demonstrated that the energy error distribution functions produced by the $r_{0}$ method are wider, more skewed and have more extended tails than those produce by $r_{opt}$.
In principle, depending on the magnitude of these differences, they could affect the determination of spectral features, specially if the spectral index is varying rapidly over a narrow energy interval such as, for example, the ankle region and beyond.

In order to assess the potential effects of using either technique for the reconstruction of a structured spectrum with rapid changes as a function of energy, we use the following semi-analytical example. An idealized sectionally continuous spectrum is assumed, that resembles the main spectral structures above $10^{18}$ eV: the ankle, the GZK flux suppression \cite{GZK-Greisen,GZK-Zatsepin} and a smooth low energy cut-off reflecting the discreteness of the surface array. The two latter suppressions in the spectrum are represented by functions of type $\tanh()$ of the input energy, while the remaining structures are represented by power laws separated by abrupt discontinuities in the first derivative. 

In order to reproduce analytically the energy error distribution functions given in Figure \ref{fig:rec_errors} as a continuous function of energy, we fit our simulation results with an Asymmetric Generalized Gaussian function (AGG):

\begin{eqnarray}\label{eq:AGG}
P_{AGG}(y)= \begin{cases}\left(\frac{c\gamma_a}{\Gamma(1/c)}\right)
\exp\{-\gamma_l^c [-(y-\mu)]^c\}& if~ y < \mu\cr \nonumber
\left(\frac{c\gamma_a}{\Gamma(1/c)}\right) \exp\{-\gamma_r^c
[(y-\mu)]^c\}&if~y \geq \mu
\end{cases} \nonumber
\end{eqnarray}

\noindent where,

\begin{equation}
\gamma_a=\frac{1}{\sigma_l + \sigma_r} \left(
\frac{\Gamma(3/c)}{\Gamma(1/c)} \right) ^{1/2} \; \;
\gamma_l=\frac{1}{\sigma_l} \left( \frac{\Gamma(3/c)}{\Gamma(1/c)}
\right)^{1/2} \; \; \gamma_r=\frac{1}{\sigma_r} \left(
\frac{\Gamma(3/c)}{\Gamma(1/c)} \right)^{1/2} \\ \nonumber
\end{equation}

\noindent and $\sigma_l^2$ and $\sigma_r^2$ are the variances of the left and right sides respectively of the probability density function and $\Gamma(x)$ is the Gamma function. If $\sigma_l^2 = \sigma_r^2$ AGG is symmetric. Furthermore, if $\sigma_l^2 = \sigma_r^2$ and $c=2$, AGG reduces to the regular Gaussian distribution function and, for $c=1$, it represents the Laplacian distribution. 

The error functions determined previously in Section \ref{SectionGauss} have been fitted using the AGG function for the both techniques: $r_{0}=600$ m and the shower-specific $r_{opt}$. In the latter case the fit reduces very nearly to a Gaussian distribution function while, for $r_{0}$, the best simultaneous fit to the right and left $\sigma_{68}$ and $\sigma_{95}$ C.L. shown in Figure \ref{fig:rec_errors}, is obtained for $c \gtrsim 2.05$. In this way we can reproduce the asymmetries present on the error distribution functions and analytically map real energies onto reconstructed ones over the whole energy range of the spectrum. 

\begin{figure}
\centering
\subfigure{\includegraphics[width=13cm]{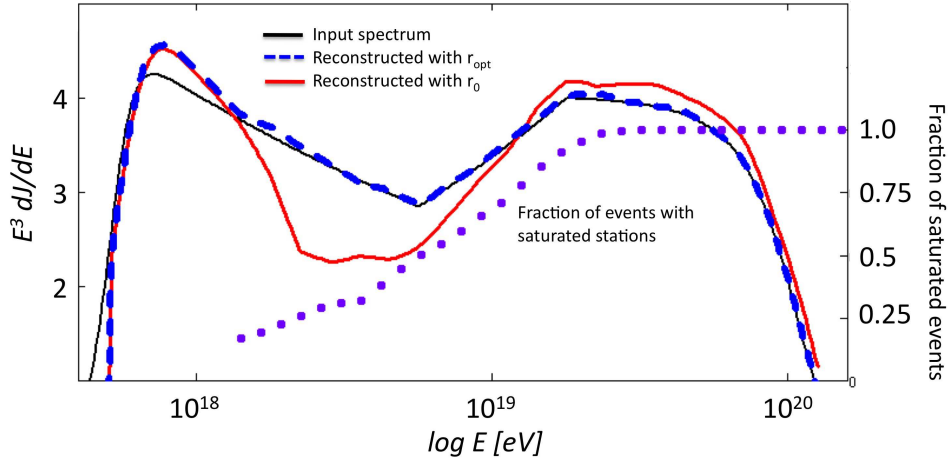}}
\caption{Input model spectrum (black/thin line) and the reconstructed spectra using $\rho(r_0)$ (red) and $\rho(r_{opt})$ (blue/dashed line) as energy estimators. The right axis shows the fraction of isotropic events between $\theta=0^{o}$ and $45^{o}$ with saturation as a function of energy (thick dotted line) for the same array with 1000 m separation.}
\label{fig:toy_spec}
\end{figure}

The results are shown in Figure \ref{fig:toy_spec}. It can be seen that, if both events with and without saturation are lumped together, the large wings associated with the error distribution functions of the $r_{0}$ approach significantly distort the spectral features. In this particular example, the ankle, is widen and shifted, while the bump and GZK suppression are shifted upwards and much more pronounced. The $r_{opt}$ approach, on the other hand, fits very tightly the original spectrum with the exception of very low energies, near the full efficiency edge, due to border effects. The $r_{0}$ approach can also give an equivalent fit, although noisier, if only events without saturation are used. However, the decrease in statistics by neglecting events with saturated detectors (cf., the fraction of events with saturation -- Figure \ref{fig:toy_spec}, right vertical axis) is so drastic that the reconstructed spectrum is only limited to a much shorter energy interval well before the GZK suppression.

In order to understand the relative magnitude of these effects, one must note that at the AGASA experiment \cite{AGASASpectrum}, for example, the systematic uncertainty in energy determination is around $18\%$. Three different sources of uncertainties combine to give this value. The first one is related with the detector, mainly its linearity ($\pm 7\%$) and response ($\pm 5\%$). Second, the uncertainties coming from the lack of knowledge of the LDF ($\pm 7\%$), the attenuation curve used ($\pm 5\%$), the shower front structure and delayed particles (which contribute $\pm 5\%$ each). Finally the relation of $\rho(r_0)$ with energy (due to the hadronic interaction model supposed, simulation codes, chemical composition etc.), introduce an uncertainty of $\pm 12\%$. In addition, they find an underestimation of $10\%$ in reconstructed energies due to energy calibration with $\rho(r_0)$, which is compensated by the overestimation due to the shower front structure and delayed particles ($5\%$ each one). We proposed that the uncertainties related to the LDF and $\rho(r_0)$ determination could be reduced by using an $r_{opt}$ determined on a shower to shower basis. However, while this may be a significant improvement, the other uncertainties would still dominate.

The Auger Observatory, a hybrid detector, reports that the largest uncertainties \cite{AugerSpectrumPRL} come from the fluorescence yield ($\pm 14\%$), the absolute calibration of FD ($\pm 10\%$), the FD reconstruction method ($\pm 10\%$). Systematic uncertainties from atmospheric aerosols, the dependence of the fluorescence spectrum on temperature and on humidity are each at the 5\% level. These uncertainties added in quadrature give a total uncertainty of $22\%$ in fluorescence energy determination. Therefore, in addition to the fact that the method proposed here does not affect directly hybrid energy reconstruction because of the improved accuracy in the determination of the core position, the total uncertainty in the spectrum determination for hybrid experiments is widely dominated by FD uncertainties.

\section{Conclusions}

The primary CR energy is generally estimated in surface arrays by interpolating the
lateral distribution function of particles in the shower front at ground level at a fixed optimum
distance $r_{0}$ from the shower core. This parameter is assumed to be predominantly dependent
on the detector separation distance for a given layout geometry and, therefore, is considered
as a constant for a given array. 

In this work we propose an algorithm to evaluate an equivalent, but shower-to-shower optimal
distance, which we call $r_{opt}$. We have performed a thorough analysis of the dependence
of $r_{opt}$ on energy and zenith angle, and demonstrate that, contrary to reference \cite{Newton}, 
these are not negligible factors.
In fact, not taking into account an event-specific $r_{opt}$, produce wider error distribution
functions that can even affect the reconstruction of a highly structured, rapidly varying spectrum.
The shower-to-shower $r_{opt}$ approach, on the other hand, is an unbiassed estimator of the CR primary energy, which produce also narrower, symmetric, almost Gaussian error distribution functions for energy reconstruction. Those properties of $r_{opt}$ can additionally lead 
to much more reliable spectral reconstruction. The differences emerging from the two procedures, $r_{0}$ vs. $r_{opt}$, when applied to spectral reconstruction may have astrophysical implications, specially in the coming era of improved precision. 

An important aspect of the $r_{opt}$ approach is that it has the additional 
advantage of allowing the same unified treatment for events with and without saturated detectors;
something that, in the $r_{0}$ approach is generally not possible, requiring either the selection of 
events through quality cuts, or the separate reconstruction with different techniques of the two 
types of events. Since the fraction of events presenting saturation is a rapidly increasing function
of energy, the later greatly reduces the effective energy range for spectral reconstruction in almost 
all practical situations. 

For practical application to real experiments, however, a proper calibration curve should be deduced specifically for $r_{opt}$, which would further optimize it as an energy estimator.

\section{Acknowledgments}

All the authors have greatly benefited from their participation in 
the Pierre Auger Collaboration. G. Ros thanks the Comunidad de Madrid for a F.P.I. fellowship and
the ALFA-EC funds, in the framework of the HELEN project. 
G. Medina Tanco also acknowledges the support of the HELEN project. This work
is partially supported by Spanish MINISTERIO DE EDUCACI\'ON Y CIENCIA under the 
projects FPA2006-12184-C02-02, CAM/UAH2005/071, CM-CCG06-UAH/ESP-0397, CSD2007-00042, 
FPA2006-12184-C02-01 and CM-UCM 910600, and Mexican PAPIIT-UNAM through grants IN115707 
and IN115607 and CONACyT through grants 46999-F and 57772. G. Ros also thanks to the 
Inst. de Ciencias Nucleares, UNAM, for its hospitality during several stays.

\end{document}